\documentclass[aps,pre,twocolumn,10pt,floatfix]{revtex4-1}

\usepackage{amsmath,bm,mathtools}
\usepackage{amssymb}
\usepackage{graphicx}
\usepackage[breaklinks]{hyperref}
\usepackage{xcolor}
\usepackage[utf8]{inputenc}
\usepackage{tikz}
\usepackage{braket}
\usepackage{braket}
\newtheorem{theorem}{Theorem}
\usepackage{dsfont}
\usepackage[normalem]{ulem}

\hypersetup{
    colorlinks,
    linkcolor={red!80!black},
    citecolor={green!70!black},
    urlcolor={blue!80!black}
}

\usepackage{hyperref}

\usepackage[]{natbib}

\PassOptionsToPackage{linktocpage}{hyperref}

\DeclareMathOperator{\erfc}{erfc}
\DeclareMathOperator{\sign}{sign}

\newcommand*{\diff}{\mathop{}\!\mathrm{d}}
\newcommand*\Diff[1]{\mathop{}\!\mathrm{d}^#1}
\newcommand*{\dgau}{\mathop{}\!\mathrm{D}}

\graphicspath{{../figura1_PRE/}{../figura2_PRE/}{../figura3_PRE/}}

\newtheorem{CSP}{Constraint satisfaction problem}

\def\mean#1{\overline{#1}}

\begin{document}

\title{Statistical learning theory of structured data}

\date{\today}
\author{Mauro Pastore}
\email{mauro.pastore@unimi.it}
\affiliation{Dipartimento di Fisica, Universit\`a degli Studi di Milano}
\affiliation{INFN, Via Celoria 16, I-20133 Milan, Italy}
\author{Pietro Rotondo}
\email{pietro.rotondo@mi.infn.it}
\affiliation{Dipartimento di Fisica, Universit\`a degli Studi di Milano}
\affiliation{INFN, Via Celoria 16, I-20133 Milan, Italy}
\author{Vittorio Erba}
\email{vittorio.erba@unimi.it}
\author{Marco Gherardi}
\email{marco.gherardi@mi.infn.it}
\affiliation{Dipartimento di Fisica, Universit\`a degli Studi di Milano}
\affiliation{INFN, Via Celoria 16, I-20133 Milan, Italy}

\begin{abstract}

The traditional approach of statistical physics to supervised learning routinely assumes 
unrealistic generative models for the data:
usually inputs are independent random variables, uncorrelated with their labels.
Only recently, statistical physicists started to explore more complex forms of data,
such as equally-labelled points lying on (possibly low dimensional) object manifolds.
Here we provide a bridge between this recently-established research area and the framework of statistical learning theory, 
a branch of mathematics devoted to inference in machine learning.
%
The overarching motivation is the inadequacy of the classic rigorous results in explaining
the remarkable generalization properties of deep learning.
We propose a way to integrate physical models of data into statistical learning theory,
and address, with both combinatorial and statistical mechanics methods, the computation of the 
Vapnik-Chervonenkis entropy,
which counts the number of different binary classifications compatible with the loss class.
As a proof of concept,
we focus on kernel machines and on two simple realizations of data structure
introduced in recent physics literature:
$k$-dimensional simplexes with prescribed geometric relations
and spherical manifolds (equivalent to margin classification).
Entropy, contrary to what happens for unstructured data, is nonmonotonic in the sample size,
in contrast with the rigorous bounds.
Moreover, data structure induces a novel transition beyond the storage capacity, which we
advocate as a proxy of the nonmonotonicity, and ultimately a cue of low generalization error.
The identification of a synaptic volume vanishing at the transition
allows a quantification of the impact of data structure within replica theory,
applicable in cases where combinatorial methods are not available,
as we demonstrate for margin learning.
\end{abstract}

\maketitle


\section{Introduction}

The idea of investigating machine learning within the tools provided by the statistical physics of disordered system is more than thirty years old, starting with the seminal papers by Amit, Gutfreund and Somplinsky \cite{PhysRevA.32.1007,PhysRevLett.55.1530} on the Hopfield model, and with Gardner's replica analysis of the Perceptron architecture \cite{Gardner:1987,GardnerDerrida:1988}.
Many of the results produced in this field have been obtained under the restrictive
and unrealistic hypothesis that the inputs of the training set were 
independent identically distributed random variables with no correlation with their labels.
Only quite recently, physicists working in this field are starting to probe the impact of more
realistic generative models of synthetic data on the available theoretical frameworks.
Sompolinsky and collaborators investigated the problem of the linear classification of perceptual manifolds \cite{ChungLeeSompolinsky:2018, ChungLeeSompolinsky:2016} and provided a first quantitative measurement of the ability to support the classification of object manifolds in deep neural networks \cite{Cohen2020}.
M\'ezard suggested that hierarchical architectures with hidden layers naturally emerge in the context of Hopfield models, assuming that the training patterns are structured as superpositions of a given set of random features \cite{Mezard:2017},
a common property of empirical data \cite{Mazzolini:2018:PRX,Mazzolini:2018:PRE}.
Zdeborov\'a and collaborators provided exact results for the generalization error within the replica approach for two different scenarios of synthetic data: random features and the hidden manifold model \cite{Goldt:2019, gerace2020generalisation}. 
One of the motivations behind these choices is the observation that many machine learning 
datasets, or their representations within deep networks, lie on the surface of low dimensional manifolds, as also verified often in practice by measuring their so-called \emph{intrinsic dimension} \cite{Cohen2020,Erba2019,AnsuiniLaio:2019,2017FaccoRodriguezEtAlEstimatingTheIntrinsicDimensionOfDatasetsByAMinimalNeighborhoodInformation,erba2020random}.
%
%
More in general, a generative model with a factorized
joint probability distribution of the inputs and their corresponding labels
is expected to be unrealistic, with respect to the
benchmark datasets commonly used in machine learning (e.g., MNIST, CIFAR-10, or Imagenet).
Intuitively, one expects there to be a notion of
similarity among inputs that constrains similar inputs to have the same label.
This regularity is expected to be related to the problem of generalization, i.e.,
the ability of a classifier to correctly classify inputs beyond the data set used for training.
%

%
The results obtained in the statistical physics framework address the typical case performance. 
In contrast, statistical learning theory (SLT) \cite{vapnik2013nature},
a successful mathematical framework in the theory of machine learning,
follows the tradition of computer science of establishing worst-case bounds.
This difference in scope made it difficult, for physicists and computer scientists alike, to work towards inter-disciplinary results,
and few examples of cross-fertilization are found in the literature.
Statistical learning theory is the branch of mathematics and computer science that studies inference,
or the problem of generating models starting from data \cite{Bousquet2004}.
It provides formal definitions for words like ``generalisation'' or ``overfitting'', 
and it is ultimately designed to evaluate the performance of learning algorithms. 
As such, it represents the ideal framework to study the problem of generalisation in deep learning.
Unfortunately, in spite of its elegance,
the insight it provides into the impressive generalization abilities of present deep learning models is poor.
The main product of the theory in this setting is a set of upper bounds
on the generalization error (which roughly counts the average number of errors made on the test set).
These upper bounds in many cases turn out to be too loose to be useful
\cite{ZhangBengio2017,MartinMahoney2017,Bottou2015, Cohn1992}.

The main drawback of this class of bounds is generally recognized to be their being
distribution independent, meaning that they hold for any probability distribution over inputs and labels of the data set,
and for all models of the chosen hypothesis class.
Substantial effort is being put, within statistical learning theory, to overcome these shortcomings
and formulate rigorous data-dependent results
\cite{Bottou2015,Antos2003,Lugosi2001,Shawe-Taylor1998}.
The Vapnik-Chervonenkis (VC) entropy is a way to establish distribution dependent, and hopefully tighter, bounds to the generalisation error \cite{Bousquet2004}. 
Informally, the VC entropy measures the number of different ways a given class of functions can classify the inputs of the training set. 
Unfortunately it is usually very difficult to compute explicitly.
Kernel architectures represent a notable exception. 
Their VC entropy has been evaluated analytically in a remarkable paper by Cover long ago \cite{cover1965}, 
under very mild hypotheses on the probability distribution of the inputs. 
The explicit calculation shows, however, that knowing the VC entropy does not improve significantly the standard bound obtained using the growth function \cite{788640}. 
In fact, both quantities scale logarithmically with the size of the training set, and depend linearly on the VC dimension,
a well known measure of model complexity.

Our goal here is to show
how the concept of data structure, as it is emerging in the 
physics literature, can be addressed within statistical learning theory,
thereby providing a bridge between the two viewpoints.
This bridge immediately allows a quantification of the generalization capabilities
of simple hypothesis classes, which shows how severely loose the classic rigorous bounds in SLT are.
Concretely, we investigate the finite-size and asymptotic behavior of the VC entropy of kernel machines by using both combinatorial and replica techniques.
While replica theory is well established in the statistical mechanics of neural networks,
combinatorial tools, though certainly not foreign to statistical mechanics 
\cite{McCoy:BOOK,CaraccioloDiGioacchino:2018,CaraccioloSportiello:2002},
have been developed only very recently for what concerns the role of data structure in machine learning
\cite{Rotondo:2020:PRR,ChungLeeSompolinsky:2018}.
We concentrate on two simple models of data structure, or ``object manifolds'':
(i) $k$-dimensional simplexes with prescribed geometric relations and (ii) spherical manifolds, which are equivalent to classify 
unstructured data points with margin (and are related to support vector machines \cite{Cortes1995}).
These models are not new, and have already received attention for their being general enough to provide
insight, but simple enough to allow full analytical treatment.

%
The manuscript is organised as follows: 
in Sec.~\ref{SLT} we review the main definitions and basic results of statistical learning theory.
In Sec.~\ref{VCkernel} we recall Cover's combinatorial result on the VC entropy of kernel architectures.
In Sec.~\ref{datastructure&SLT} we discuss how data structure can be taken into account in SLT,
and define the two synthetic data ensembles we use in the following.
In Sec.~\ref{combinatorialres} we first recall
the extension of Cover's combinatorial technique to structured data that was introduced in \cite{Rotondo:2020:PRR},
and then we establish the asymptotic behavior of the VC entropy for the first ensemble (simplexes).
After introducing an asymptotic method based on analytic combinatorics in Secs.~\ref{sec:analytic_combinatorics}
and \ref{sec:combinatorial_unstructured},
we use it to show that the VC entropy is non-monotonic in the load in Sec.~\ref{sec:non_monotonicity}.
In Sec.~\ref{sec:satisfiability_transition} we describe a satisfiability transition that is brought about by data structure
\cite{companion1},
and we further analyse it in Secs.~\ref{section:generic_k} and \ref{sec:finite_size}.
In section \ref{replicares}, we introduce a synaptic volume that monitors the behavior of the VC entropy in the thermodynamic limit.
This is particularly useful for those object manifolds for which the combinatorial method is not yet available,
which includes our second data ensemble (spherical manifolds).
We present calculations for the annealed (in Sec.~\ref{sec:annealed}) and quenched (in Sec.~\ref{sec:quenched}) averages
(in the replica symmetric and one-step replica symmetry breaking ans\"atze)
for the case of $2$-dimensional simplexes, and for spherical manifolds in Sec.~\ref{sec:spheres}.

\section{Taking data structure into account in statistical learning theory}
\label{datastructureSLT}

\subsection{Basic results in statistical learning theory}
\label{SLT}

In this section we recall the basic facts of SLT, mostly following the exposition of \cite{Bousquet2004}.
We restrict to binary classification problems, in which the goal is to find a function $g$ mapping the input space $\mathcal X$
to the output space $\mathcal Y = \{+1,-1\}$.
Each pair $Z^\mu = ( X^{\mu}, Y^\mu)$ (with $\mu = 1, \dots, p$) in the training set 
$Z_p = (Z^1,\dots, Z^p)$ is drawn 
by the unknown joint probability distribution $P_{\mathcal X, \mathcal Y} (X,Y)$. 
A map $g$ between the set of inputs $X_p=$ and $\{+1,-1\}$ is called a dichotomy of $X_p$.
The criterion to choose $g$ is the minimization of the risk
\begin{equation}
R(g) = \braket {\mathds 1_{g(X) \neq Y}}_P,
\end{equation}
which is the probability of error.
Ideally, we should look for
$\inf_g R(g)$ over all the possible $g$'s.
Since $P$ is unknown, the best we can do is to consider the empirical risk
\begin{equation}
R_p (g) = \frac{1}{p} \sum_{\mu = 1}^p  \mathds 1_{g(X^\mu) \neq Y^\mu},
\end{equation}
and limit the search within a specific hypothesis class $\mathcal G$ to prevent overfitting.
A dichotomy $g\in\mathcal G$ is called realizable.
The output of a learning algorithm is a function $g_p$ that depends on the data $Z_p$.
The goodness of the choice of $g_p$ can be measured by its generalization error $\epsilon_\mathrm{gen}(g_p)$,
where
\begin{equation}
\epsilon_{\mathrm{gen}}(g) = R(g) - R_p(g).
\end{equation}
Notice that $\epsilon_\mathrm{gen}(g)\leq 1$.
In practice, $R_p$ is evaluated on the training set and $R$ is estimated on a test set
\cite{Mehta:2019:PR}.
One of the primary goals of SLT
is to establish rigorous bounds on the generalization error.

A complementary description of risk minimization within a class $\mathcal G$
is given through the definition of the loss class $\mathcal L$:
\begin{equation}
\label{lossclass}
\mathcal L = \left\{\ell_g :  (X,Y) \mapsto \mathds 1_{g(X) \neq Y},\, g \in \mathcal G \right\}.
\end{equation}
To each $g \in \mathcal G$, we associate a function $\ell_{g}$ such that $\ell_{g}((x,y))=1$ if $g(x)\neq y$, and is zero otherwise.
While elements of $\mathcal G$ take values in $\{+1,-1\}$,
those of $\mathcal L$ have range $\{0,1\}$.
$\ell_{g_p}$ can be used to count the number of errors made on the training set
by the function $g_p$.
Given a loss class $\mathcal L$, we can consider its projection on the sample $Z_p$, by defining
\begin{equation}
\label{eq:projected_lossclass}
\mathcal L_{Z_p} = \left\{(\ell (Z^1), \ell (Z^2), \dots, \ell (Z^p))\,: \, \ell \in \mathcal L  \right\}.
\end{equation}
This is the set of all possible ways that a function in $\mathcal G$
can correctly or incorrectly classify each sample in $Z_p$.
Importantly, $\mathcal L_{Z_p}$ can be interpreted as the set of all
classifications of the points in $X_p$ that can be realized by the model, i.e.,
the set of all $(Y^1,\ldots,Y^p)$ such that there exists a $g\in\mathcal{G}$
such that $Y^\mu=g(X^\mu)$ for all $\mu$.
This representation reveals a useful bijection between $\mathcal L_{Z_p}$
and the set of realizable dichotomies.

A key quantity in SLT is the
Vapnik-Chervonenkis (VC) entropy $\mathcal H_{\mathcal L} (Z_p)$, which
measures the size of $\mathcal L_{Z_p}$:
\begin{equation}
\mathcal H_{\mathcal L} (Z_p) = \log \left|\mathcal L_{Z_p}\right|.
\end{equation}
By virtue of the bijection discussed above, valid when $\mathcal Y=\{+1,-1\}$,
$\mathcal H_{\mathcal L} (Z_p)$ can be defined equivalently as
\begin{equation}
\mathcal H_{\mathcal L} (Z_p) = \log \mathcal N_\mathcal{G}(X_p),
\label{VCentropy}
\end{equation}
where $\mathcal N_{\mathcal G}(X_p)$ is the number of dichotomies of the set $X_p$ realizable by $\mathcal G$.
The VC entropy controls a rigorous upper bound to the generalization error:
\begin{theorem}
\label{Th1}
For any $\delta > 0$, with probability at least $1-\delta$, 
\begin{equation}
\forall g \in \mathcal G,\ \epsilon_{\mathrm{gen}}(g) \leq 2 \sqrt{2 \frac{\mathcal H_{\mathcal L} (2 p) + \log\frac{2}{\delta}}{p}}\,,
\end{equation}
where the annealed VC entropy $\mathcal H_{\mathcal L} (p)$ is defined as:
\begin{equation}
\mathcal H_{\mathcal L} (p) = \log \braket{\mathcal N_{\mathcal L} (Z_p)}
\label{annVCentropy}
\end{equation}
and $\braket{\cdot}$ is the average over the joint probability distribution $P_{\mathcal X,\mathcal Y}$ of the training set.
\end{theorem}

Unfortunately, direct computation of the VC entropy is unfeasible in most cases. 
For this reason, a main goal of SLT is to construct more tractable upper bounds to the VC entropy. 
%
The classic example is based on the Vapnik-Chervonenkis dimension, 
which is a scalar metric of the expressivity of a given hypothesis class $\mathcal G$.
More formally, the VC dimension $d_{\mathrm{VC}}$ of a class $\mathcal G$ is the largest integer such that
there exists at least one set of $d_{\mathrm{VC}}$ inputs $X_{d_\mathrm{VC}}$ such that
\begin{equation}
\mathcal N_\mathcal{G}(X_{d_\mathrm{VC}}) = 2^{d_\mathrm{VC}}
\end{equation}
(i.e., the class $\mathcal G$ realizes all possible dichotomies of the inputs).
With this definition, it can be proved that
\begin{equation}
\label{entropybound}
\mathcal H_{\mathcal L} (p) \leq d_{\mathrm{VC}} \log \left( \frac{e p}{d_{\mathrm{VC}}}\right).
\end{equation}
Hence, a corollary of Theorem~\ref{Th1} is the well-known upper bound first obtained by Vapnik:
if the class $\mathcal G$ has finite VC dimension $d_{\mathrm{VC}}$, then, with probability at least $1-\delta$,
\begin{equation}
\label{VCbound}
\forall g \in \mathcal G,\ \epsilon_{\mathrm{gen}}(g) \leq 2 \sqrt{2 \frac{d_{\mathrm{VC}} \log \left( \frac{2ep}{d_{\mathrm{VC}}}\right) + \log\frac{2}{\delta}}{p}}.
\end{equation}
A crucial property of this elegant result is its being distribution independent,
meaning that the bound is uniform in the function $g$, and
does not depend on the particular problem at hand.
Owing to its universality, the bound is often too loose for most practical applications~\cite{Bottou2015}.
Let us consider for instance a deep neural network with a number of weights $w = 10^6$--$10^9$. 
In this case the VC dimension is of order $d_{\mathrm{VC}} \sim w \log w$ \cite{sontag1998vc}.
When the typical size of the dataset is $p=10^4$--$10^6$, as is often the case in practice, 
is is evident that bounds such as the one in Eq.~\eqref{VCbound} do not offer any insight 
on the generalization performance of deep neural networks.  
Indeed, one of the main pursuits of contemporary SLT is to provide better results on the generalization error,
going beyond distribution independent bounds.
Several strategies have been proposed,
advocating the importance of considering data-dependent hypothesis classes~\cite{Shawe-Taylor1998}
and data-dependent measures of complexity
(such as the Rademacher complexity~\cite{BartlettMendelson:2003}, which was recently connected to the statistical mechanics
of disordered systems~\cite{AbbaraAubin:2019}),
also in relation to the original concept of VC entropy itself \cite{AnguitaGhio:2014}.

\subsection{Vapnik-Chervonenkis entropy of kernel machines}
\label{VCkernel}

As mentioned above, in most cases it is not possible to compute the VC entropy directly.
However, kernel machines are a notable exception: their VC entropy was computed
half century ago by Cover \cite{cover1965}.
Kernel architectures provide a special realization of one-hidden layer neural networks and are at the core of the idea of support vector machines. In these machines, one defines \emph{a priori} a kernel function $\phi : \mathbb{R}^n \rightarrow \mathbb R^d$, that maps $n$-dimensional inputs to a $d$-dimensional feature space. One of the simplest realizations of such maps is a quadratic polynomial kernel, such that each input $X$ is mapped on a $d(d+1)/2$-dimensional feature space via a kernel $\phi^{(2)}$ with components $\phi^{(2)}_{ij} = X_i X_j$, $\forall i \leq j$.
The map from feature space to the space of labels is realized by a linear separator:
\begin{equation}
Y = \sign(W\cdot X),
\end{equation}
where the weight vector, $W\in\mathbb{R}^n$, is the set of learnable parameters.

Cover's theorem is a function counting theorem: it computes the number of dichotomies 
$\mathcal N_\phi(X_p)$ of this function class,
the logarithm of which is the VC entropy.
It is simpler to state Cover's theorem for linear separators, 
i.e., for $d=n$ and $\phi=\mathds 1$;
the realizable dichotomies in this case are called linearly realizable.
We comment below on the extension to general $\phi$.
The key idea behind the theorem is twofold:
(i) under a weak condition on the inputs $X_p$,
the number of dichotomies $\mathcal N_{\mathds 1}(X_p)$ is a function solely of the dimension $n$ and the number of points $p$;
(ii) it is possible to write a solvable recurrence relation, in $n$ and $p$, for this function.
Following Cover's original paper, we denote the (data-independent) number of dichotomies $\mathcal N_{\mathds 1}(X_p)$ 
by $C_{n,p}$, and the corresponding VC entropy by $\mathcal{H}_{n,p}=\log C_{n,p}$.
\begin{theorem}[Cover, 1965]
\label{thm:cover}
Let $X_p$ be a set of $p$ points in $\mathbb{R}^n$.
If the points are in general position, i.e., if the points in $X'$
are linearly independent for all subsets $X'\subseteq X_p$ such that $\left| X'\right|\leq n$, then
$\mathcal N_{\mathds 1}(X_p)=C_{n,p}$, where
\begin{equation}
\label{eq:Cnp_k1}
C_{n,p}=2\sum_{j=0}^{n-1} \binom{p-1}{j}.
\end{equation}
\end{theorem}
The proof of Theorem \ref{thm:cover} is based on a simple recurrence relation for $C_{n,p}$:
\begin{equation}
\label{eq:recurrence_k1}
C_{n,p+1}=C_{n,p}+C_{n-1,p},
\end{equation}
with boundary conditions
\begin{equation}
\label{eq:cover_BC_k1}
C_{n\geq 1,1}=2, \quad C_{0,p}=0.
\end{equation}
Equation \eqref{eq:recurrence_k1} states that adding the $(p+1)$th point $X$ to $X_p$
increases the number of dichotomies by $C_{n-1,p}$, which is the number of dichotomies of $X_p$
that are realizable by a vector $W$ such that $W\cdot X=0$.
Cover actually proved a more general statement.
Informally, if one maps all elements of $X_p$ by the non-linear kernel function $\phi$
from $\mathbb{R}^n$ to $\mathbb{R}^{d}$ with $d$ larger than $n$,
then, under mild assumptions on $\phi$, Eq.~\eqref{eq:Cnp_k1} holds with $d$ in place of $n$.

It is straightforward to see how the storage capacity defined in statistical mechanics, $\alpha_\mathrm{c}$ (recall that $\alpha=p/n$),
can be obtained from $C_{n,p}$.
The number of dichotomies is a combinatorial quantity, 
and is expected to scale exponentially in $n$, at least for small $\alpha$.
Thus, an intensive quantity can be defined by normalizing $C_{n,p}$ with the total number of dichotomies of $p$ points.
The fraction of dichotomies $c_{n,p}\equiv C_{n,p}/2^p$ is bounded, $0\leq c_{n,p}\leq 1$,
and has a non-trivial thermodynamic limit $c_\infty(\alpha)$.
The thermodynamic limit is defined by taking both $n,p\to\infty$, with fixed $\alpha=p/n$.
It is not hard to see directly from Eq.~\eqref{eq:Cnp_k1} that
\begin{equation}
\label{eq:c_theta}
c_\infty(\alpha) = \theta\left(\alpha_\mathrm{c} - \alpha \right),
\end{equation}
with $\alpha_\mathrm{c}=2$.
The expression in Eq.~\eqref{eq:c_theta} takes the value $1$ for $\alpha<\alpha_\mathrm{c}$, 
the value $0$ for $\alpha>\alpha_\mathrm{c}$,
and the value $1/2$ for $\alpha=\alpha_\mathrm{c}$
($\theta$ is the Heaviside step function).
Qualitatively, $c_{n,\alpha n}$ as a function of $\alpha$ is a decreasing sigmoid, 
which is steeper for larger values of $n$
(see Fig.~\ref{fig:fractionVSnumber}a).
This allows the definition of a notion of capacity at finite dimension $n$,
as the value $\tilde\alpha_\mathrm{c}(n)$ such that $c_{n, \tilde\alpha_\mathrm{c}(n) n} = 1/2$, or
\begin{equation}
\label{eq:alphac_finitesize}
C_{n, \tilde\alpha_\mathrm{c}(n) n} = 2^{p-1}.
\end{equation}
Another notable value of $p$ can be read off of $c_{n,p}$: it is the Vapnik-Chervonenkis
dimension $d_\mathrm{VC}$, equal to the maximum $p$ such that $c_{n,p}=1$.
For a linear separator, $d_\mathrm{VC}=n$.
Notice that one cannot use the asymptotic form Eq.~\eqref{eq:c_theta} to this aim,
since the thermodynamic limit pushes $c_{n,\alpha n}$ to $1$ for all values of $\alpha$
up to $\alpha_\mathrm{c}$.
Notice that Eq.~\eqref{eq:Cnp_k1} implies that the VC entropy grows asymptotically as
$\mathcal H_{n,p} \sim (n-1)\log p$
for large number of inputs $p$
(see Sec.~\ref{sec:combinatorial_unstructured} for a derivation).
This is the same behavior as that obtained by bounding the VC entropy
as in Eq. \eqref{entropybound}.


Two remarks can be made, concerning the generality of Cover's theorem.
First,
the general position is a rather weak condition.
For instance, we mention three examples of distributions of the points 
$\xi_\mu\in X_p$ under which the general position holds with probability $1$:
(i) $\xi^\mu\in X_p$ are i.i.d.~variables with the uniform measure on the sphere $S^{n-1}$;
(ii) $\xi^\mu\in X_p$ are i.i.d.~variables with marginal probability distribution $P(\xi)$, and the support of $P$ is $\mathbb R^n$;
(iii) the coordinates of each $\xi^\mu\in X_p$ are i.i.d.~variables, with discrete probability distribution
$p(x) = (1+m)/2 \delta_{x,1} + (1-m)/2 \delta_{x,-1}$, for any $m\in [-1,1]$.
Clearly, there are trivial ways to violate general position:
for instance, if the probability distribution of (i) or (ii) above is conditioned to assigning the same value
to a fixed subset of size $k < n$ of the coordinates of all inputs.
Then Cover's theorem still applies in the subspace, with $n-k$ in place of $n$.

Second,
the condition that $\phi$ must satisfy for the theorem to apply to the kernel machine specified by $\phi$
is essentially that the vectors $\phi(\xi^\mu)$ must be in general position in the feature space $\mathbb R^d$.
This again is a very mild condition.
Starting with a set of inputs $X_p$ in general position in the original $n$-dimensional space,
most interesting mappings satisfy the condition. 
This includes polynomial kernels, but also more complex functions, such as those of the form $\phi_i (\xi)= g\left(\sum_j W_{ij} \xi_j \right)$, where $g$ is an activation function (e.g., $\mathrm{ReLU}$ or $\tanh$) and $W$ is any rectangular random matrix. 
The latter case is relevant for the theory of extreme learning machines \cite{HUANG2006489}.

\subsection{Constrained models of structured data}
\label{datastructure&SLT}

The discussion above suggests that, in order to go beyond the prediction of Cover's theorem,
one needs a way of introducing statistical dependence between 
the inputs $X_p$ and their labels $Y_p = (Y^1,\ldots,Y^p)$.
This reflects a simple observation that can be made on empirical datasets of images:
similar inputs tend to be classified similarly.
For instance, one expects that there exists an (unknown) set of transformations on an input image $X$,
possibly including some translations, dilations, and rotations, that leave the classification of $X$ invariant.
Such intuition agrees with the concepts, put forward in neuroscience and gaining momentum in physics,
of invariant recognition
(the similar neural representation of the same object in different conditions) and
object manifolds (sets of input stimuli giving rise to the same neural representation)
\cite{Cohen2020,ChungLeeSompolinsky:2018,ChungLeeSompolinsky:2016,AnselmiLeibo:2016,Seung:2000}.

Integrating data structure within the framework of statistical mechanics is relatively straightforward
and usually follows two steps:
(i) define a generative model for the data,
given in terms of a nonfactorized joint probability distribution $P(X_p,Y_p)$;
(ii) compute averages over the measure $P$ (the ``disorder'');
this is what was done for instance in \cite{Borra:2019,ChungLeeSompolinsky:2018, ChungLeeSompolinsky:2018, gerace2020generalisation}.
How to best address data dependence in the SLT formalism, instead, is a debated issue.
Here we follow a simple strategy inspired by recent literature in statistical physics:
we change the input space $\mathcal X$.
Each input $X^\mu$ is now an object manifold, i.e., a 
(possibly countably or uncountably infinite) set of points that, by definition, are be classified coherently.

We focus on two simple realizations of data structure,
the first motivated by the availability of analytical results
and the second motivated by its connection to the well-known framework of margin learning.
\paragraph{Simplex learning} --- 
Inputs are ``multiplets'' of $k$ points with fixed geometric interrelations.
The input set is $X_p= \{X^\mu\}_{\mu=1,\ldots,p}$, 
where each $X^\mu = \{\xi_a^\mu\}_{a=1,\dots, k}$ is a set of $k$ points 
on the unit $(n-1)$-sphere, $\xi_a^\mu \in S^{n-1}$. 
The $k(k-1)/2$ overlaps within each multiplet are fixed: $\xi_a^\mu \cdot \xi_b^\mu = \rho_{ab}$ for all $\mu = 1,\dots, p$. 
We assume the uniform probability measure on each point $\xi_a^\mu$, conditioned on the
constraint on the overlaps \cite{Rotondo:2020:PRR}.
The usual unconstrained ensemble is recovered for $k=1$, or at any $k$ if $\rho_{ab} =1$ for all $a,b$.
The name ``simplex'' is justified by the fact that, since linear classification is a projective problem,
if $Y=g(X)$ for each $X$ in a set of points $X^\mu$, then $Y=g(X)$ for all $X$ in the convex hull of $X^\mu$.
The input space $\mathcal X_\mathrm{S}\left(\{\rho_{ab}\}\right)$ depends on $k$ and $\rho_{ab}$, and is the set of all
multiplets with the given constraints.
\paragraph{Margin learning} --- 
Given a kernel machine with kernel $\phi:\mathbb R^n\to \mathbb R^d$
and inputs $X\in\mathcal X=\mathbb R^n$, learning with margin $\kappa$ 
is defined by the class $\mathcal G(\kappa)$ of all functions
\begin{equation}
g_\kappa(X)=
\begin{cases}
+1 \quad W\cdot \phi (X) > \kappa\\
-1  \quad W\cdot \phi (X) <  -\kappa.
\end{cases}
\end{equation}
Cases falling within the margin $(-\kappa, \kappa)$ can be defined with a third value, for instance $0$,
or left undefined.
Hence, the corresponding loss class projected on a sample $(X_p, Y_p)$, 
Eqs.~\eqref{lossclass} and \eqref{eq:projected_lossclass},
contains all the dichotomies of $X_p$ that can be realized by an element of $\mathcal G_\mathrm{M}(\kappa)$.
An alternative representation of margin learning can be given via the definition of appropriate object manifolds.
In fact, linear separation of points with margin $\kappa$ is equivalent to zero-margin 
linear separation of spherical object manifolds with radius $\kappa$~\cite{ChungLeeSompolinsky:2018}.
Thus, $Y^\mu=g_\kappa(X^\mu)$ for all $\mu$ if and only if
$Y^\mu=g_0(Q^\mu)$ for all $\mu$ and all $Q^\mu$ such that $\left| Q^\mu-\phi(X^\mu)\right|^2<\kappa^2$.
The input space $\mathcal X_\mathrm{M}(\kappa)$ is the set of the preimages, via $\phi$, 
of all spheres of radius $\kappa$ in $\mathbb R^d$.
%
%
Note that, while margin learning has a natural description in terms of the original space $\mathcal X=\mathbb R^n$,
through the hypothesis class $\mathcal G_\mathrm{M}(\kappa)$,
simplex learning does not have such a straightforward representation,
and is defined directly by means of the object space $\mathcal X^\mathrm{S}\left(\{\rho_{ab}\}\right)$.
The VC entropy for margin learning, $\mathcal H_\kappa$, can be bounded from above 
by means of the VC dimension $d_\mathrm{VC}(\kappa)$:
\begin{equation}
\mathcal H_\kappa \leq d_{\mathrm{VC}} (\kappa) \log p, \quad p > d_{\mathrm{VC}}(\kappa).
\end{equation}
In turn, an upper bound of the VC dimension exists for points
lying on the $d$-dimensional sphere of radius $R$~\cite{788640}: 
\begin{equation}
d_{\mathrm{VC}}(\kappa) \leq \min \left[ \frac{R^2}{\kappa^2},d \right].
\end{equation}
The standard bound is therefore again logarithmic in the sample size $p$.
In the following, we set out to investigate the behavior of the VC entropy of kernel machines 
for the two data structures defined above,
in order to quantify how loose these logarithmic upper bounds are.
We do so by means of two complementary approaches:
the combinatorial framework and the theory of disordered systems.

\section{Combinatorial approach}
\label{combinatorialres}

Very recently the combinatorial approach introduced by Cover
was extended to formulate a mean field theory of simplex learning \cite{Rotondo:2020:PRR}.
In this section we focus on this model of data structure.
The definition we have given above of simplex learning specifies the ensemble of the sets $X_p$.
It remains to define the hypothesis class $\mathcal G_\mathrm{M}(\{\rho_{ab}\})$.
This is straightforward: one starts from the class $\mathcal G$ of linear separators in $\mathbb R^n$
and restricts it to the class $\hat{\mathcal G}(\{\rho_{ab}\})$ of
those functions $h\in \mathcal G$ that assign the same label to all points in each multiplet $X^\mu$
(i.e., those that are constant on each multiplet).
Then the restricted hypothesis class is defined as
\begin{equation*}
\begin{split}
\mathcal G_\mathrm{M}(\{\rho_{ab}\})=& \Big\{ g :
\exists h\in \hat{\mathcal G}(\{\rho_{ab}\}) \;\mathrm{s.t.}\; \\
&\phantom{\Big\{ g :}
\forall X^\mu\in X_p,
g(X^\mu) = h(\xi\in X^\mu)
\Big\}.
\end{split}
\end{equation*}
The functions in $\hat{\mathcal G}(\{\rho_{ab}\})$ are called admissible.
The mean-field combinatorial theory allows the computation of the average 
$\left<\mathcal N_{\mathds 1}(X_p)\right>_{X_p}$, i.e.,
the average number of admissible dichotomies of simplexes that can be realized linearly.
We will still denote this number with $C_{n,p}$, although it depends on the parameters $k$ and $\{\rho_{ab}\}$ of the ensemble.

The quantities $C_{n,p}$ satisfy a recurrence relation
\begin{equation}
\label{eq:recurrence_k}
C_{n,p+1}=\sum_{l=0}^k \theta^k_l C_{n-l,p},
\end{equation}
where the constant coefficients $\theta^k_l$ are fixed in turn by the recurrence relation
\begin{equation}
\label{eq:theta_recursion}
\theta^k_l=\psi_k \theta^{k-1}_l + \left(1- \psi_k \right) \theta^{k-1}_{l-1},
\end{equation}
with boundary conditions 
\begin{equation}
\begin{split}
&\theta^1_0=\theta^1_1=1\\
&\theta^k_{-1}=\theta^k_{k+1}=0.
\end{split}
\end{equation}
The boundary conditions for Eq.~\eqref{eq:recurrence_k} are difficult to express precisely for generic $k$.
Here we will assume the boundary conditions in Eq.~\eqref{eq:cover_BC_k1} for all $k$.
This approximation is expected to have a negligible effect for the asymptotic analysis
presented in the following;
we checked the validity of this approximation numerically for the first non-trivial cases $k=2$ and $k=3$. 
Each coefficient $\theta_l^k$ in Eq.~(\ref{eq:recurrence_k}) depends on $k-1$ numbers 
$\{\psi_m\}_{m=2,\ldots,k}$, with $0\leq\psi_m\leq 1$,
having the following geometric-probabilistic interpretation.
Let $w\in S^{n-1}$ be a random vector with the flat measure on the unit sphere.
Consider any multiplet $X^\mu$, and a subset
$X'\subseteq X^\mu$ of $m\leq k$ points.
Then $\psi_m$ is the symmetrized probability that the scalar product $w\cdot\xi$
has the same sign for all $\xi\in X'$, conditioned on it having the same sign
for all $\xi\in X'\setminus\{\xi_\star\}$:
\begin{equation*}
\psi_m = 2 \left<\mathrm{P}\left[ (w\cdot \xi_\star) > 0 \;|\; (w\cdot \xi) > 0\; 
\forall \xi\in X'\setminus\{\xi_\star\} \right]\right>_{\mathrm{sym}},
\end{equation*}
where the symmetrization $\left<\cdot\right>_\mathrm{sym}$ is performed by averaging over all subsets $X'$
and over all choices of $\xi_\star\in X'$.
These quantities can be expressed in terms of the overlaps $\rho_{ab}$,
e.g.,
\begin{equation}
\psi_2(\rho)=\frac{2}{\pi}\arctan \sqrt{\frac{1+\rho}{1-\rho}}.
\end{equation}
(More information on $\psi_m$ can be found in \cite{Rotondo:2020:PRR}).

The notion of storage capacity $\alpha_\mathrm{c}$ can be defined for structured data,
similarly to Cover's unstructured case, and coherently with the thermodynamic
limit addressed in statistical mechanical computations.
The combinatorial theory yields
\begin{equation}
\label{eq:capacity_k}
\alpha_\mathrm{c}(k)=\left(
k-\frac{1}{2}-\sum_{l=2}^k \psi_l \right)^{-1}.
\end{equation}

\subsection{Asymptotic analysis via analytic combinatorics}
\label{sec:analytic_combinatorics}

\begin{figure}[tb]
\includegraphics[scale=1.1]{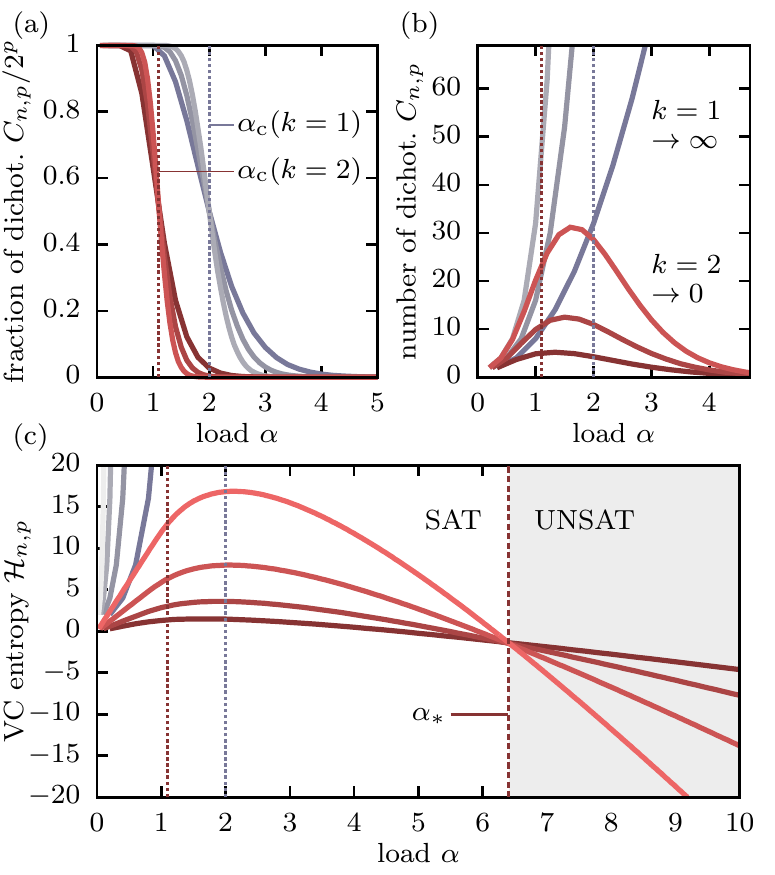}
\caption{
While the fraction of admissible dichotomies (a) has qualitatively similar behavior
for unstructured (grey curves, $k=1$) and structured (red curves, $k=2$) data,
the absolute number of dichotomies (b) has different limit behaviors.
As a consequence, the VC entropy (c) diverges to $+\infty$ for unstructured data
and to $-\infty$ for structured data.
Curves of the VC entropy at different values of $n$ intersect, for large $n$,
at the same critical value $\alpha_*$ of the load.
Vertical dotted lines in all panels are the storage capacities.
The dashed line in (c) is the transition caused by data structure.
[$n=5,10,20$ in (a), $n=3,4,5$ in (b), $n=5,10,20,40$ in (c).]
}
\label{fig:fractionVSnumber}
\end{figure}

In the case of unstructured data, we know that
the growth of $C_{n,p}$ as a function of $p$
is exponential up to the capacity $p_\mathrm{c}=2n$
and sub-exponential afterwards.
Due to this change of behavior, the fraction of linearly realizable dichotomies,
$c_{n,p}=C_{n,p}/2^p$,
has a discontinuous transition from $1$ to $0$ in the thermodynamic limit
(see Fig.~\ref{fig:fractionVSnumber}a).
What is the asymptotic growth rate of $C_{n,p}$?
This question can be answered by inspecting the explicit solution Eq.~\eqref{eq:Cnp_k1}.
However, we construct a different method here, based on the techniques of analytic combinatorics.
Our method has the crucial advantage
of being applicable to cases where
(i) the solution $C_{n,p}$ is not known explicitly, and (ii) the recurrence equation
is given implicitly, as a relation between its coefficients.

Let $g_n(z)$ be the ordinary generating function of $C_{n,p}$ with respect to the variable $p$:
\begin{equation}
\label{eq:gn_generating_function}
g_n(z)=\sum_{p=1}^\infty C_{n,p} z^p.
\end{equation}
Formally, the coefficient $C_{n,p}$ can be obtained by derivation as
\begin{equation}
C_{n,p} = \left. \frac{1}{p!} \frac{\mathrm d^p}{\mathrm d z^p} g_n(z) \right|_{z=0}.
\end{equation}
When it is unfeasible to compute the $p$-th derivative explicitly,
one can extract information on the asymptotic behavior of $C_{n,p}$ for large $p$
by means of analytic techniques
(see for instance \cite{flajolet2009analytic}).

Whenever the generating function Eq.~(\ref{eq:gn_generating_function}) is a rational function analytic in $z=0$,
it admits a partial fraction expansion
\begin{equation}
\label{eq:g_Q_sum}
g_n(z) = Q_n(z) + \sum_{s}\sum_{r=1}^{r_s} \frac{a_{s,r}}{(z-z_s)^r},
\end{equation}
where $Q_n$ is a polynomial,
$s$ ranges over the poles of $g_n$, and $r_s$ is the multiplicity of the pole $s$.
Then, the asymptotic form of the coefficients of $g_n(z)$ 
can be read off the series expansion of $(z-z_s)^{-r}$:
\begin{equation}
\label{eq:negative_binomial_series}
\left(z-z_s\right)^{-r} = \frac{(-1)^r}{z_s^r} \sum_{p=0}^\infty \binom{p+r-1}{r-1} z_s^{-p} z^p.
\end{equation}
By substituting (\ref{eq:negative_binomial_series}) in Eq.~\eqref{eq:g_Q_sum} one obtains $r_s$ different
contributions for each pole $s$.
The overall leading term corresponds to the
dominant singularity $z_0$ of $g_n(z)$, i.e., the one with smallest modulus $|z_0|$.
This is due to the term $z_s^{-p}$ in (\ref{eq:negative_binomial_series}) that
suppresses the sub-dominant poles exponentially.
Among the contributions due to $z_0$, the leading one is that with $r=r_s$,
because the binomial coefficient in (\ref{eq:negative_binomial_series}) is a polynomial of degree $r-1$ in $p$.
Putting it all together,
if the dominant singularity is a pole of order $r$, then
\begin{equation}
\label{eq:asymptotic_analysis}
C_{n,p} \sim R z_0^{-p-r} \binom{p+r-1}{r-1},
\end{equation}
where the constant $R$ can be obtained by factoring out the singularity:
\begin{equation}
\label{eq:singularity_R}
R = \lim_{z\to z_0} (z_0-z)^r g_n(z).
\end{equation}
Equation~(\ref{eq:asymptotic_analysis}) shows that
if $|z_0|<1$ (respectively, $>1$), $C_{n,p}$ increases (respectively, decreases) exponentially with $p$ at fixed $n$;
if $|z_0|=1$ then the asymptotic behavior is polynomial (of order $r-1$).

In simple cases, when it is possible to obtain $g_n(z)$ in closed form,
this method can be applied straightforwardly.
However, this set up allows to probe the asymptotics of $C_{n,p}$
even in more complicated scenarios, where $g_n(z)$ cannot be solved for explicitly,
or when even the recurrence relation for $g_n(z)$ is not specified completely.
Section \ref{section:generic_k} shows how to tackle this more general problem.
Before that, we consider the simpler cases $k=1$ and $k=2$.

\subsection{Asymptotics for unstructured data}
\label{sec:combinatorial_unstructured}

As a ``warm-up exercise'', we use the combinatorial method described above
to explore the asymptotics of $C_{n,p}$ in the well-understood unstructured case.

By multiplying both sides of Eq.~(\ref{eq:recurrence_k1}) by $z^p$ and summing over $p$
one obtains
\begin{equation}
\frac{1}{z} g_n(z) - 2 = g_n(z) + g_{n-1}(z),
\end{equation}
where the constant term $2$ comes from the initial condition \eqref{eq:cover_BC_k1}.
It is useful to rewrite the equation as
\begin{equation}
\label{eq:gn_recurrence_k1}
g_n(z)=\frac{z}{1-z}\left[ g_{n-1}(z) + 2 \right].
\end{equation}
The boundary condition is $g_0(z)=0$, due to every $C_{0,p}$ being zero.
The relation (\ref{eq:gn_recurrence_k1}) is a linear (non homogeneous)
first-order recurrence with constant coefficients,
whose solution is
\begin{equation}
\label{eq:gn_k1}
g_n(z)=\frac{2z}{2z-1}\left[\left(\frac{z}{1-z}\right)^n-1\right].
\end{equation}
Equation~(\ref{eq:gn_k1}) shows that $g_n(z)$ has a single pole
at $z_0=1$, of order $n$, with finite part $R=2$.
Therefore, the corresponding asymptotic form has no exponential factor,
and is purely polynomial:
\begin{equation}
\label{eq:asymptotic_k1}
C_{n,p}\sim 2 \binom{p+n-1}{n-1} = \frac{2}{(n-1)!} p^{n-1} + O\left(p^{n-2}\right).
\end{equation}
Note that
the right-hand side of Eq.~(\ref{eq:gn_k1}) has a removable discontinuity in $z_1=1/2$, 
where the apparent pole in the first term gets canceled by a zero in the numerator
(the term in square brackets).
The corresponding exponential asymptotic growth, $2^p$, is present in $C_{n,p}$ only transiently, 
for $p<n$.

\subsection{Non-monotonicity of the VC entropy}
\label{sec:non_monotonicity}

The behavior of $C_{n,p}$, and therefore of the VC entropy, changes dramatically
when data structure is present, already in the simplest case where
the training data are structured as pairs of points, i.e., $k=2$.
Figure \ref{fig:fractionVSnumber} shows the fraction of dichotomies, $C_{n,p}/2^p$,
and the number of dichotomies, $C_{n,p}$, as functions of $\alpha$
for increasing values of the dimension $n$, for $k=1$ and $k=2$ with $\rho=0.3$.
The fraction of dichotomies is qualitatively similar in the two scenarios,
the only apparent difference being the expected decrease in the storage capacity.
A remarkable divergence appears instead in the asymptotic behavior of $C_{n,p}$.
The absolute number of dichotomies is non-monotonic for simplex learning
already in the simplest nondegenerate case $k=2$ with $\rho<1$.
What is also evident in Fig.~\ref{fig:fractionVSnumber}b is the fact that the storage capacity $\alpha_\mathrm{c}(k)$
does not pinpoint any qualitatively special point for the unnormalized $C_{n,p}$,
and therefore for the VC entropy.

Since the two-point case $k=2$ is the simplest case where the nonmonotonicity of the VC entropy arises,
we work it out in detail, before showing the general $k$-point case below.
The geometry of the problem is fixed by the single quantity $\psi_2$.
The recurrence equation reads
\begin{equation}
\label{eq:recurrence_k2}
C_{n,p+1} = 
\psi_2 C_{n,p} + C_{n-1,p} + \left(1-\psi_2\right) C_{n-2,p},
\end{equation}
with boundary conditions
$C_{0,p}=0, C_{n,1}=2\{1-[1-\psi_2(d)]\delta_{n,1}\}$.
In order to simplify the computations, we will use the same boundary conditions
as for $k=1$, i.e., $C_{0,p}=0$ and $C_{n\geq 1, 1}=2$.
This approximation has negligible effects in the large-$n$ limit \cite{Rotondo:2020:PRR}.

Equation (\ref{eq:recurrence_k2}) fixes the recurrence relation satisfied by
the generating function $g_n(z)$:
\begin{equation}
g_n(z)=\frac{z}{1-\psi_2 z} \left[ g_{n-1}(z) + (1-\psi_2) g_{n-2}(z) + 2 \right],
\end{equation}
with boundary condition $g_{n\leq 0}(z)=0$.
The solution, which can be found by means of the characteristic polynomial method,
reads
\begin{equation*}
\begin{split}
g_n(z)= &\left[\frac{z-\sqrt{\Delta(z)}}{2(1-\psi_2 z)}\right]^n
\frac{z}{2z-1}\left(1+z\frac{2\psi_2-3}{\sqrt{\Delta(z)}} \right) \\
+&\left[\frac{z+\sqrt{\Delta(z)}}{2(1-\psi_2 z)}\right]^n
\frac{z}{2z-1}\left(1-z\frac{2\psi_2-3}{\sqrt{\Delta(z)}} \right)
-\frac{2z}{2z-1},
\end{split}
\end{equation*}
where $\Delta(z)=z[4(1-\psi_2)+z(1-2\psi_2)^2]$.
The explicit solution has a pole of order $n$
in $z_0=1/\psi_2$, with finite part
\begin{equation}
R=2 \psi_2^{-2n}.
\end{equation}
Similarly to the unstructured case, the singularity in $z=1/2$ is removable,
which signals that the initial exponential increase of the number of dichotomies
must be superseded eventually by the asymptotic behavior due to $z_0$.
Altogether, the large-$p$ form of $C_{n,p}$ is
\begin{equation}
\label{eq:asymptotic_k2}
C_{n,p}\sim 2 \binom{p+n-1}{n-1} \psi_2^{p-n}.
\end{equation}

The crucial difference between the results for $k=1$, Eq.~\eqref{eq:asymptotic_k1},
and $k=2$, Eq.~\eqref{eq:asymptotic_k2}, lies in the fact that
while the first is asymptotically increasing, the second is exponentially decreasing
whenever $\psi_2<1$, i.e., when the two partner points are distinct.
Observe that $C_{n,p}$ always increases for small $p$;
this is a consequence of the fact that the Vapnik-Chervonenkis dimension
of a linear classifier in $n$ dimensions is $d_\mathrm{VC}=n$,
therefore all dichotomies of $kp$ points can be realized 
when $p\leq n/k$, meaning that $C_{n,p\leq n/k} = 2^p$.
The decreasing asymptotic form then proves that
the Vapnik-Chervonenkis entropy $\mathcal{H}_{n,p}$
is non-monotonic in $p$ (and therefore in $\alpha$) for fixed $n$.
Intuitively,
the non-monotonicity is due to the competition of two opposing effects.
On one hand, the addition of a new pair of points $\{\xi,\bar\xi\}$ to a set of $p$ existing pairs
entails a combinatorial increase in the total number of linearly-realizable dichotomies.
On the other hand, some of the $C_{n,p}$ admissible dichotomies
can become invalid if they are realizable only by hyperplanes intersecting the segment
connecting $\xi$ and $\bar\xi$.

\subsection{Emergence of a data-driven satisfiability transition}
\label{sec:satisfiability_transition}

A non-trivial consequence of the non-monotonic VC entropy
can be observed in Fig.~\ref{fig:fractionVSnumber}c.
Consider the VC entropy $\mathcal{H}_{n,\alpha n}$ as a function of $\alpha$.
The curves $\mathcal{H}_{n,\alpha n}$ at different values of $n$
intersect each other roughly around the same point $\alpha_*$.
More precisely, if $\mathcal{H}_{n,\alpha n}$ and $\mathcal{H}_{n-1,\alpha (n-1)}$
intersect at $\alpha_*(n)$, then $\alpha_*=\lim_{n\to\infty}\alpha_*(n)$.
This empirical observation can be clarified analytically.

As a function of the load $\alpha=p/n$, Eq.~(\ref{eq:asymptotic_k2}) becomes
\begin{equation}
\label{eq:asymptotic_alpha_k2}
C_{n,\alpha n}  \sim C(\alpha;n) \equiv
2\frac{\Gamma\left(\alpha n + n\right)}{\Gamma(n)\Gamma\left(\alpha n+1\right)}\psi_2^{(\alpha-1)n}
\end{equation}
($\Gamma$ is the Euler gamma function),
or $\mathcal{H}_{n,\alpha n}  \sim \mathcal{H}(\alpha;n)$ with
\begin{equation}
\label{eq:asymptotic_entropy_alpha_k2}
\mathcal{H}(\alpha;n) \equiv
\log\left[2\frac{\Gamma\left(\alpha n + n\right)}{\Gamma(n)\Gamma\left(\alpha n+1\right)}\right]+
(\alpha-1)n\log \psi_2.
\end{equation}
In the non-degenerate case (whenever $\psi_2<1$) the second term
in \eqref{eq:asymptotic_entropy_alpha_k2} is negative for $\alpha>1$,
while the first term is always positive.
This competition gives rise to a transition at $\alpha=\alpha_*>1$, where the asymptotic limit 
of the VC entropy changes:
\begin{equation}
\lim_{n\to\infty} \mathcal{H}(\alpha;n) = \left\{
\begin{matrix}
-\infty & \alpha<\alpha_*\phantom{.}\\
\infty & \alpha>\alpha_*.
\end{matrix}
\right.
\end{equation}
The transition point is pinpointed by the condition
\begin{equation}
\lim_{n\to\infty}\frac{\mathrm d}{\mathrm d n} \mathcal{H}(\alpha_*;n) = 0.
\end{equation}
With $\mathcal{H}(\alpha;n)$ given by Eq.~(\ref{eq:asymptotic_entropy_alpha_k2}), the condition reads
\begin{equation}
\begin{split}
\lim_{n\to\infty}
&\left[(\alpha_*-1)\log\psi_2+(\alpha_*+1)\Psi(\alpha_* n+n) \right. \\
&\left.-\Psi(n)-\alpha_*\Psi(\alpha_* n+1)\right] = 0,
\end{split}
\end{equation}
where $\Psi(z)\equiv\partial_z\log \Gamma(z)$ is the poly-gamma function,
whose asymptotic behavior is $\Psi(z)=\log(z)+O(1/z)$.
Sending $n$ to infinity then gives the transcendental equation
\begin{equation}
\label{eq:alpha_equation_k2}
(\alpha_*+1)\log(\alpha_*+1)-\alpha_*\log\alpha_*+(\alpha_*-1)\log\psi_2 = 0,
\end{equation}
which has two solutions: $\alpha_*$ is the larger.
As a function of $\psi_2$, the transition point $\alpha_*$ has limits
\begin{equation}
\label{eq:alpha_limits}
\begin{split}
\lim_{\psi_2\to 0} \alpha_* &= 1 \\
\lim_{\psi_2\to 1} \alpha_* &= \infty.
\end{split}
\end{equation}
As expected, when $\psi_2$ goes to $1$, the problem reduces to that of classifying unstructured data,
and the transition runs to infinity.

The phase transition at $\alpha_*$ can be rationalized as the SAT-UNSAT transition 
of a random constraint satisfaction problem (CSP).
First, we recall that the storage capacity $\alpha_\mathrm{c}$ itself
corresponds to the transition between the satisfiable and the unsatisfiable phase
of an appropriate satisfiability problem.
The CSP relevant to $\alpha_\mathrm{c}$ can be stated as follows:
\begin{CSP}
\label{problem:1}
Given a set of $kn$ input-label pairs $\{\xi_a^\mu, \sigma^\mu\}$
(with $a=1,\ldots,k$ and $\mu=1,\ldots,p$), find
a vector $w$ such that $\sign(w \cdot \xi_a^\mu) = \sigma^\mu$ for all $\mu$ and $a$.
\end{CSP}
The input data of this problem satisfies the admissibility constraints by construction.
A corresponding random constraint satisfaction problem (rCSP) is
an ensmble of CSPs, specified by a probability measure on the input data.
The rCSP is in the SAT (respectively UNSAT) phase when the satisfiability problem
admits a solution with probability one (respectively zero) in the thermodynamic limit.
The storage capacity \eqref{eq:capacity_k}
marks the transition between the SAT and the UNSAT phases of the rCSP
corresponding to problem \ref{problem:1} with the probability measure
of simplex learning described in Sec.~\ref{datastructure&SLT} %

A different problem can be constructed by moving the admissibility property
from the definition of the input data to the conditions defining the solution:
\begin{CSP}
\label{problem:2}
Given a set of $kn$ input points $\{\xi_a^\mu\}$,
(with $a=1,\ldots,k$ and $\mu=1,\ldots,p$), find
a set of labels $\{\sigma^\mu\}$ and
a vector $w$ such that $\sign(w \cdot \xi_a^\mu) = \sigma^\mu$ for all $\mu$ and $a$.
\end{CSP}
Notice that this problem is trivially satisfiable for unstructured data,
i.e., it is satisfied by almost all vectors $w$ when the constraint of admissibility
is irrelevant (i.e., when $k=1$).
A solution to problem \ref{problem:2} is given by specifying
an admissible dichotomy $\{\sigma^\mu\}$ and a vector $w$.
In this framework, the VC entropy counts the (logarithm of the) number of distinct dichotomies $\{\sigma^\mu\}$
that can appear in such a solution.
This means that the corresponding  rCSP is in the UNSAT phase when $\mathcal{H}(\alpha;n)\to -\infty$
and in the SAT phase otherwise.
%

\subsection{Transition point for generic \texorpdfstring{$k$}{k}}
\label{section:generic_k}

Now we address the more general case where the number of partners
in a multiplet is $k$.
The generating function $g_n(z)$ satisfies the recurrence equation
\begin{equation}
\label{eq:recurrence_gn_k}
g_n(z)=\frac{z}{1-z \theta_0^k} \left[ 2+ \sum_{l=1}^k \theta^k_l g_{n-l}(z)\right],
\end{equation}
as can be obtained from Eq.~\eqref{eq:recurrence_k}.
Solving for $g_n(z)$ from Eqs.~(\ref{eq:recurrence_gn_k}) and (\ref{eq:theta_recursion}) would be hopeless.
However, the asymptotic analysis discussed above only needs three pieces of information
about $g_n(z)$, namely (i) the location $z_0$ of the dominant singularity, (ii) its order $r$, and (iii) its finite part $R$.
These can be extracted from the recurrence relations without solving them.

The right-hand side of Eq.~(\ref{eq:recurrence_gn_k}) has a singularity
in $z=1/\theta^k_0$.
The boundary condition is $g_{n\leq0}(z)=0$,
therefore the first non-zero function is $g_1(z)=2\sigma(z)$,
where
\begin{equation}
\label{eq:singularity_sigma}
\sigma(z)=\frac{z}{1-z\theta^k_0} 
\end{equation}
encapsulates the singularity.
Since the number of terms in the sum in Eq.~(\ref{eq:recurrence_gn_k}) is finite,
no other singularity can appear at finite $n$.
Therefore
\begin{equation}
z_0=\frac{1}{\theta^k_0}.
\end{equation}
Now consider one iteration of Eq.~\eqref{eq:recurrence_gn_k}: the singularity with largest order
in the right-hand side comes from $g_{n-1}(z)$, and the singular term
gets multiplied by $\theta^k_1 \sigma(z)$.
Indeed, it is easy to see by induction that the leading term $\hat g_n(z)$ in the Laurent
expansion of $g_n(z)$ around $z_0$ is
\begin{equation}
\label{eq:gn_solution_k}
\hat g_n(z) = 2 \left(\theta^k_1\right)^{n-1} \sigma(z)^n.
\end{equation}
Therefore, the order of the singularity is $r=n$. 
The constant $R$ [Eq.~\eqref{eq:singularity_R}]
can be obtained by multiplying Eq.~(\ref{eq:gn_solution_k}) by $(1/\theta^k_0-z)^n$
and evaluating it at $z=1/\theta^k_0$:
\begin{equation}
R=2\left(\theta^k_1\right)^{n-1}\left(\theta^k_0\right)^{-2n}.
\end{equation}
Finally, the asymptotic behavior of $C_{n,p}$ is
\begin{equation}
C_{n,p}\sim 2 \binom{p+n-1}{n+1} \left(\theta^k_1\right)^{n-1} \left(\theta^k_0\right)^{p-n},
\end{equation}
from which one readily obtains the asymptotic form $C(\alpha;n)$ for the number of dichotomies,
\begin{equation}
\label{eq:C_alpha_n}
C(\alpha;n) = 2 \frac{\Gamma(\alpha n+n)}{\Gamma(n)\Gamma(\alpha n+1)} 
\left(\theta^k_1\right)^{n-1} \left(\theta^k_0\right)^{(\alpha-1)n},
\end{equation}
and the corresponding one for the VC entropy,
\begin{equation}
\label{eq:H_alpha_n}
\begin{split}
\mathcal{H}(\alpha; n) = &\log\left[2 \frac{\Gamma(\alpha n+n)}{\Gamma(n)\Gamma(\alpha n+1)}\right]\\
&+(n-1)\log \theta^k_1 + (\alpha-1)n\log\theta^k_0.
\end{split}
\end{equation}

As above, the existence of a critical value $\alpha_*$ can be established by
finding the zeros of the derivative of $\mathcal{H}(\alpha;n)$ with respect to $n$,
in the large-$n$ limit.
One finds
\begin{equation}
\label{eq:alpha_equation_k}
(\alpha_*+1)\log(\alpha_*+1)-\alpha\log\alpha_*
+(\alpha_*-1)\log\theta^k_0 +\log\theta^k_1 = 0.
\end{equation}

The two coefficients $\theta^k_0$ and $\theta^k_1$ can be obtained
from Eq.~(\ref{eq:theta_recursion}) as functions of the $\psi$'s.
By solving the recurrence equation, specialized to $l=0$, one has
\begin{equation}
\label{eq:theta_k_0}
\theta^k_0=\prod_{m=2}^k \psi_m.
\end{equation}
Then, by substituting expression \eqref{eq:theta_k_0} into Eq.~(\ref{eq:theta_recursion}) 
with $l=1$, one obtains the recurrence relation
\begin{equation}
\theta^k_1=\psi_k \theta^{k-1}_1+(1-\psi_k)\prod_{m=2}^{k-1}\psi_m,
\end{equation}
with boundary condition $\theta^1_1=1$.
The solution is
\begin{equation}
\label{eq:theta_k_1}
\theta^k_1=\left(2-k+ \sum_{m=2}^k \frac{1}{\psi_m} \right) \prod_{m=2}^k \psi_m.
\end{equation}
Specializing to $k=3$, for instance, yields
\begin{equation}
\begin{split}
\theta^3_0 &= \psi_3 \psi_2\\
\theta^3_1 &= \psi_3 + \psi_2 - \psi_3\psi_2.
\end{split}
\end{equation}
Because of the way $\theta^k_0$ and $\theta^k_1$ are constructed via
the geometric quantities $\psi_m\in[0,1]$, they are not independent.
The range of $\theta^k_0$ is $[0,1]$, as can be seen from Eq.~(\ref{eq:theta_k_0}).
The sup and inf of $\theta^k_1$ at fixed $\theta^k_0$ can be obtained
by considering the two extremal cases
\begin{equation}
\begin{split}
\mathrm{(i)}\: &\left\{\psi_m\right\}_m=\left\{1,\ldots,1,\theta^k_0,1,\ldots,1\right\},\\
\mathrm{(ii)}\: &\left\{\psi_m\right\}_m= \{ \left(\theta^k_0\right)^{1/(k-1)}, \ldots, \left(\theta^k_0\right)^{1/(k-1)} \}.
\end{split}
\end{equation}
The fact that the evaluation on the two extremal cases gives the appropriate bounds is not obvious:
it can be proved by induction using Lagrange's theorem for constrained optimization 
(taking care to consider the boundary of the domain as well); see Appendix \ref{app:bounds_theta}.
From (i) and (ii) respectively one gets
\begin{equation}
\label{eq:inf_sup_theta}
\begin{split}
\mathrm{(i)}\: &\sup \theta^k_1=1,\\
\mathrm{(ii)}\: &\inf \theta^k_1=(k-1) \left(\theta^k_0\right)^{1-\frac{1}{k-1}} + (2-k)\theta^k_0.
\end{split}
\end{equation}
The inf is monotonically decreasing with $k$; therefore, by letting $k\to\infty$
one obtains a global lower bound independent of $k$:
\begin{equation}
\label{eq:bound_theta}
\theta^k_1 > \theta^\infty_1 = \theta^k_0 \left[ 1-\log \theta^k_0 \right].
\end{equation}
The upper bound (i) is already $k$-independent.

\begin{figure}[tb]
\includegraphics[scale=1.12]{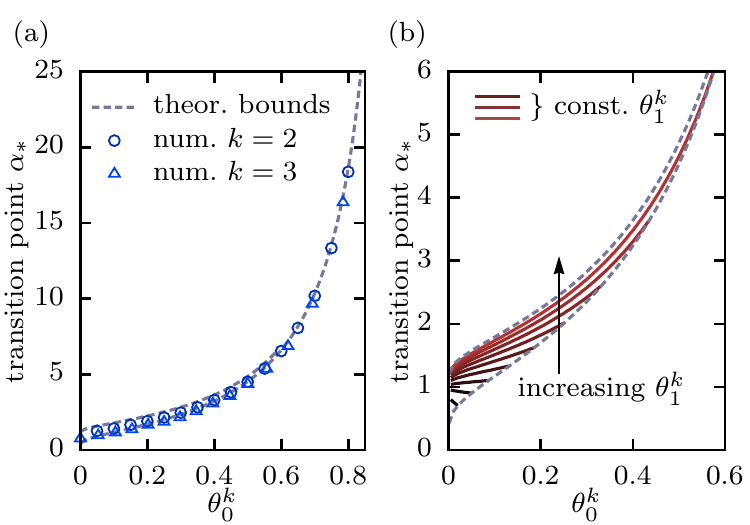}
\caption{
(a) Numerical estimates of $\alpha_*$ at varying $\theta^k_0$ for two different geometries:
$k=2$ (where $\theta^2_0$ is just $\psi_2$) and $k=3$. In the latter case we fix $\{\rho_{ab}\}$
by requiring that the three points in the simplex form an equilateral triangle of varying sizes.
(b)
Theoretical results (red curves) for $\alpha_*$ as a function of $\theta^k_0$ for
increasing values of $\theta^k_1$, within its allowed range given by
Eqs.~\eqref{eq:inf_sup_theta} and \eqref{eq:bound_theta}.
Dashed lines in both panels are the $k$-independent upper and lower bounds for $\alpha_*$.
}
\label{fig:2}
\end{figure}

Figure \ref{fig:2} summarizes the results concerning the value of $\alpha_*$ for generic $k$.
It also shows a comparison with numerical results obtained for $k=3$,
(with $\{\rho_{ab}\}$ given by the equilateral geometry).
The theoretical bounds in the figure (dashed lines) are obtained by
substituting the $k$-independent bounds above into Eq.~(\ref{eq:alpha_equation_k}).
%

We point out that there are two sources of approximation in the computations above,
namely (i) the modified boundary conditions,
and (ii) the perturbative nature of the asymptotic analysis.
Concerning (i), we remark that the numerical results were obtained
by using the correct boundary conditions.
However, using the modified conditions does not change the numerical
results appreciably.
The small discrepancies apparent in the Fig.~\ref{fig:2} are therefore due almost entirely to (ii).

\subsection{Finite-size scaling at the critical point}
\label{sec:finite_size}

In the vicinity of the transition point $\alpha_*$,
the quantity $C(\alpha;n)$ satisfies finite-size scaling,
as happens for other random satisfiability problems \cite{Kirkpatrick1994, Leone2001}.
In this section we compute the scaling form and its critical exponents.

Let us define a scaling variable $y$ as $n$ times the reduced load $(\alpha-\alpha_*)/\alpha_*$ around $\alpha_*$:
\begin{equation}
y = n \frac{\alpha-\alpha_*}{\alpha_*}.
\end{equation}
By inserting $\alpha = \alpha_* y/n + \alpha_*$ in Eq.~(\ref{eq:C_alpha_n}),
and using the asymptotic expansion of the $\Gamma$ function,
\begin{equation}
\Gamma(x) = e^{x\log x - x} \left[\sqrt{2\pi} x^{-1/2} + \mathrm{O}\left(x^{-3/2}\right)\right],
\end{equation}
one obtains in the large-$n$ limit
\begin{equation*}
C(\alpha;n) = 
e^{nA+B}\left[  \frac{\sqrt{2/\pi}}{\sqrt{\alpha_*(1+\alpha_*)}} n^{-1/2}  + \mathrm{O}\left(n^{-3/2}\right) \right],
\end{equation*}
with
\begin{equation*}
\begin{split}
A &= (\alpha_*+1)\log(\alpha_*+1)-\alpha_*\log\alpha_* \\
&\phantom{=}+(\alpha_*-1)\log\theta^k_0 +\log\theta^k_1,\\
B &= - \log\theta^k_1 + \alpha_* y \log(\alpha_*+1) - \alpha_* y \log\alpha_* + \alpha_* y \log \theta^k_0.
\end{split}
\end{equation*}
The linear term $nA$ in the exponential vanishes by Eq.~(\ref{eq:alpha_equation_k}).
Hence,
\begin{equation}
\begin{split}
C(\alpha;n) = n^{-1/2} \frac{1}{\theta^k_1} \frac{\sqrt{2/\pi}}{\sqrt{\alpha_*(1+\alpha_*)}} \left(\frac{\alpha_*+1}{\alpha_*} \theta^k_0 \right)^{\alpha_* y}\\
\times\left[ 1 + \mathrm{O}\left(n^{-3/2}\right) \right],
\end{split}
\end{equation}
which shows that in the thermodynamic limit $C(\alpha;n)$ obeys the scaling form
\begin{equation}
\label{eq:scaling_form}
C(\alpha;n) = n^{-1/2} F\left( \frac{\alpha-\alpha_*}{\alpha_*} n  \right)
\end{equation}
with the exponential scaling function
\begin{equation}
\label{eq:exp_scaling_function}
F(y) = \frac{1}{\theta^k_1} \frac{\sqrt{2/\pi}}{\sqrt{\alpha_*(1+\alpha_*)}} \left(\frac{\alpha_*+1}{\alpha_*} \theta^k_0 \right)^{\alpha_* y}.
\end{equation}
Equation~\eqref{eq:scaling_form} shows that, within the approximation of our asymptotic analysis,
the number of dichotomies satisfies the finite-size scaling form
\begin{equation}
C_{n,\alpha n} \sim n^{-\beta/\nu} F\left( \frac{\alpha-\alpha_*}{\alpha_*} n^{1/\nu}\right) 
\end{equation}
(where $F$ is regular),
with critical exponents
\begin{equation}
\beta = 1/2,\quad \nu = 1.
\end{equation}

Let $h(\alpha)$ be the VC entropy density in the thermodynamic limit:
\begin{equation}
h(\alpha) = \lim_{n\to\infty}\frac{1}{n} \mathcal{H}(\alpha;n).
\end{equation}
The condition $h(\alpha)=0$, satisfied by $\alpha_*$, can be written from Eq.~\eqref{eq:exp_scaling_function} as
\begin{equation}
\label{eq:alpha_star_condition}
\left(\alpha-\alpha_*\right)\log\left(\frac{\alpha_*+1}{\alpha_*}\theta^k_0\right) = 0.
\end{equation}
Curiously, Eq.~(\ref{eq:alpha_star_condition}) is satisfied identically in $\alpha$
if $\alpha_*=\theta^k_0/(1-\theta^k_0)$.
By plugging this value of $\alpha_*$ into Eq.~(\ref{eq:alpha_equation_k}), one obtains the
simple condition $\theta^k_1=\theta^k_0(1-\theta^k_0)$.
For data structure with $\theta^k_0$ and $\theta^k_1$ satisfying this relation, 
one therefore expects that $C(\alpha;n)$ is constant in $\alpha$ in the large-$n$ limit;
equivalently, the VC entropy will be approximately independent of the load,
$\mathcal{H}_{n,p}\sim\mathcal{H}_n$.


\section{Replica approach}
\label{replicares}

The discussion in the foregoing sections shows that 
(i) the VC entropy has nonmonotonic behavior for simplex learning,
(ii) the hallmark of the nonmonotonicity is  the existence of a phase transition,
and (iii) the transition can be framed as the SAT-UNSAT transition of a constraint satisfaction problem, 
which is different from the one that defines the storage capacity. 
Since it is often challenging to deal with the combinatorics of complex data structures, our goal in this section
is to identify an appropriate synaptic volume that provides access to the transition.
Once this observable is identified, we will be able to pinpoint the existence of the phase transition without direct access to the VC entropy, in the same spirit of the original work by Gardner \cite{Gardner:1987}, by using disordered systems techniques.

We define the synaptic volume by leveraging on the definition of the CSP corresponding to the transition.
As already noted, in looking for a solution to the constraint satisfaction problem \ref{problem:2} 
(defined in Sec.~\ref{sec:satisfiability_transition}), we have 
the freedom to adjust both the synaptic weights $W$ and the outputs $\sigma$. 
This means that the outputs are promoted to be dynamical variables and should be treated at the same level of the synaptic weights. 
This suggests that the relevant synaptic volume for identifying the corresponding phase transition is the following:
\begin{equation}
\begin{aligned}
V(X_p) 
{}={}& \sum_{\{\sigma^\mu = \pm 1\}}\int \left[ \prod_{j=1}^n \diff W_j \right]\delta\! \left(\sum_{j=1}^n W_j^2 -n \right)\\
&\times \prod_{\mu=1}^{p}  \prod_{a=1}^{k}\theta\!\left(\frac{\sigma^\mu}{\sqrt{n}} \sum_{j=1}^n W_j \xi_{a,j}^\mu  \right),
\label{eq:replica_volume}
\end{aligned}
\end{equation}
where $\theta(\cdot)$ is the Heaviside theta, $\xi_{a,j}^\mu$ denotes the $j$-th component of the $a$-th element of the $\mu$-th multiplet and the weights lie on the surface of a $n$-dimensional sphere of radius $\sqrt n$
(note that this is different from the convention used in the preceding sections).
The inputs, constituting the set $X_p$, are chosen randomly according to the distribution
\begin{equation}
\begin{aligned}
\diff P(X_p) 
{}={}&\nu^{-1} \prod_{\mu=1}^p\prod_{a=1}^k  \prod_{b=1}^{a-1} \delta\!\left(\rho_{ab} - \frac{1}{n}\sum_{j=1}^n \xi^\mu_{a,j} \xi^\mu_{b,j}\right) \\
&\times \prod_{j=1}^n\left[\delta(\xi^\mu_{a,j} - 1) + \delta(\xi^\mu_{a,j} +1)\right]   \diff \xi_{a,j}^\mu \,,
\label{eq:replica_inputs}
\end{aligned}
\end{equation}
where $-1\le \rho_{ab}\le 1$ are the overlaps, $\nu$ is a normalization factor and the inputs lie on the vertices of a $n$-dimensional hypercube.

Note that data structure is implemented in Eq.~\eqref{eq:replica_volume} by asking that each point of the $\mu$th symplex be labelled by $\sigma^\mu$. Moreover, this synaptic volume differs from the ordinary Gardner volume by the integration over the labels $\sigma$, considered dynamical variables on the same foot of the weights $W$.
Intuitively, an exponential growth of $V(X)$ with $n$ at fixed load $\alpha$ means that,
in the thermodynamic limit,
at least one classification compatible with the input-label constraints can be expressed by the model;
on the contrary, when $V(X)$ decreases exponentially in $n$ then no
such classification exists for $n\to\infty$.
Thus, the logarithm of $V(X)$ is a suitable observable to assess the nonmonotonic behavior of the VC entropy for a given data structure.

We will apply replica theory to compute the averaged (over the inputs positions) logarithm of the synaptic volume defined in 
Eq.~\eqref{eq:replica_volume}, in order to identify the transition. The goal will be the evaluation of the critical value of $\alpha=p/n$ where this volume changes regime, as a function of the overlaps. In the following, we will restrict to the case $k=2$, i.e. to data organised in doublets, so that the geometry of the simplex is fully specified by a single parameter $\rho$; to lighten the notation, we will omit the index $a=1,2$, simply denoting the doublets as $(\xi,\bar{\xi})$. Using standard integral representations for the delta and theta functions, we can write the volume of interest as
\begin{equation}
\begin{aligned}
V {}={}& \sum_{\{\sigma^\mu = \pm 1\}}\int \left[ \prod_{j=1}^n \diff W_{j} \right] \int_{0}^{+\infty} \left[\prod_{\mu=1}^p \frac{\diff \lambda^\mu \diff \bar{\lambda}^\mu }{(2\pi)^2} \right]\\
&\!\!\times\int_{-\infty}^{+\infty} \left[ \prod_{\mu=1}^p \diff x^\mu \diff \bar{x}^\mu \right]\int_{-\infty}^{+\infty} \frac{\diff E}{2\pi} \, e^{iE\left(\sum_j W_{j}^2 - n\right)}\\
&\!\!\times e^{ i \sum_{\mu} x^\mu \left(\lambda^\mu - \frac{\sigma^\mu}{\sqrt{n}} \sum_{j} W_{j} \xi_j^\mu \right)+ i \sum_{\mu} \bar{x}^\mu \left(\bar{\lambda}^\mu - \frac{\sigma^\mu}{\sqrt{n}} \sum_{j} W_{j} \bar{\xi}_j^\mu \right)},
\label{eq:replica_volume_k2}
\end{aligned}
\end{equation}
where the auxiliary variable $E$ enforces the spherical constraint, while the standard integral representation of the theta function is obtained via the auxiliary variables $\lambda$, $x$.

We dedicate the following sections to the calculation of the averaged logarithm of this volume in the annealed, replica symmetric (RS) and one-step replica symmetry breaking (1RSB) approximations. The main results of this section, to which we address the reader not interested in the details, are Eq.~\eqref{eq:replica_alphaA}, \eqref{eq:replica_alphaRS} and \eqref{eq:replica_alpha1RSB}.
 
\subsection{Annealed computation}
\label{sec:annealed}

The annealed calculation is based on the substitution $\mean{\log V}\to \log \mean{V}$, so we simply need to average the volume \eqref{eq:replica_volume_k2} with respect to the input distribution (indicated by the overline); the details are reported in Appendix~\ref{app:input}. After a large-$n$ expansion and the average over the inputs, the integrals in $x$ and $\lambda$ can be solved explicitly:
\begin{equation}
\begin{aligned}
&\Biggl[\sum_{\{\sigma=\pm 1\}}\int_{0}^{+\infty}  \frac{\Diff2 \bm{\lambda} }{(2\pi)^2}  \int_{-\infty}^{+\infty} \Diff2 \bm{x} \, e^{-\frac{1}{2} \bm{x}^T \mathcal{R} \bm{x} + i \bm{x}^T \bm{\lambda} } \Biggr]^p\\
&\quad=\Biggl[2\int_{0}^{+\infty}  \frac{\Diff2 \bm{\lambda} }{(2\pi)^2}  \frac{2\pi}{\sqrt{1-\rho^2}} e^{-\frac{1}{2}\bm{\lambda}^T \mathcal{R}^{-1} \bm{\lambda}  } \Biggr]^p\\
&\quad=\Biggl[\frac{1}{2} + \frac{1}{\pi} \arcsin \rho \Biggr]^p\,,
\end{aligned}
\label{eq:replica_average_annealed}
\end{equation}
where we introduced the notation
\begin{equation}
\bm{x} = \begin{pmatrix}
x\\
\bar{x}
\end{pmatrix}\,,\quad \bm{\lambda} = \begin{pmatrix}
\lambda\\
\bar{\lambda}
\end{pmatrix}\,,\quad
\mathcal{R} = \begin{pmatrix}
1&\rho\\
\rho&1
\end{pmatrix}
\end{equation}
and we used the known formula for the quadrant probability of a bivariate normal distribution, see \cite{gupta1963}. The remaining integrals can be performed: the one over the weights is Gaussian
\begin{equation}
\int \left[ \prod_{j=1}^n \diff W_{j} \right]   e^{i E \sum_j W_{j}^2} = e^{n \left[ \frac{1}{2}\log \pi - \frac{1}{2} \log (-iE) \right]}\,,
\end{equation}
while the one over $E$ can be performed via a saddle-point method for large $n$:
\begin{equation}
\begin{aligned}
&\int_{-\infty}^{+\infty} \frac{\diff E}{2\pi}\,  e^{-in E - \frac{n}{2} \log (-iE)} \sim \frac{1}{2\sqrt{\pi n}} e^{n\left[\frac{1}{2} + \frac{ \log 2}{2}\right]}\,.
\end{aligned}
\end{equation}
Assembling everything, and ignoring inessential factors, we find
\begin{equation}
\mean{V} =  \exp \left\{n \left[ \frac{p}{n} \log \left(\frac{1}{2} + \frac{1}{\pi} \arcsin \rho \right)  + \frac{1 + \log 2\pi  }{2} \right]  \right\}\,.
\end{equation}
Defining the critical value of $\alpha=p/n$ as the one where the exponent changes sign, we find
\begin{equation}
\alpha^{A}_*(\rho) = - \frac{ 1 + \log 2\pi  }{2 \log\left(\frac{1}{2} + \frac{1}{\pi} \arcsin \rho \right)} \,.
\label{eq:replica_alphaA}
\end{equation}

\begin{figure}[tb]
\includegraphics[scale=1.12]{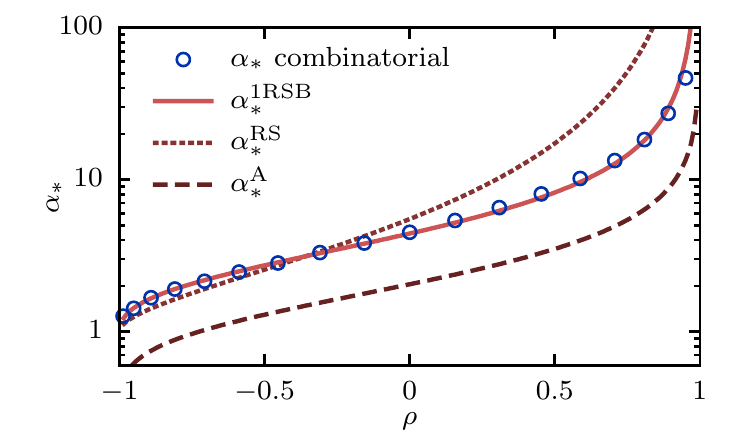}
\caption{Critical value of the load $\alpha$ as a function of the overlap $\rho$ for $k=2$ (data in pairs). Circles represent the combinatorial result, which is in agreement with numerical simulations. All the different approximation schemes used for the replica computations display the same qualitative shape. However the annealed and RS ansatz fail in reproducing quantitatively the combinatorial result. Using a 1RSB ansatz we obtain a one-parameter expression for $\alpha_\ast$ [Eq. \eqref{eq:replica_alpha1RSB}] that 
fits the combinatorial result tightly.}
\label{fig:replica_alpha}
\end{figure}

A comparison of the annealed approximation and of the result obtained with combinatorics in Eq.~\eqref{eq:alpha_equation_k2} is shown in Fig.~\ref{fig:replica_alpha}. 
Although the annealed approximation fails in reproducing quantitatively the behavior of $\alpha_{\ast} (\rho)$, 
it bounds the combinatorial result from below, and qualitatively recovers
the expected divergence for $\psi_2 \to 1$.

\subsection{Quenched computation}
\label{sec:quenched}

The quenched calculation of $\mean{\log V}$ is performed via the replica trick. First, we replicate $t$ times the volume \eqref{eq:replica_volume_k2}, obtaining
\begin{equation}
\begin{aligned}
&V^t =  \sum_{\{\sigma^\mu_a =\pm 1\}} \int  \left[\prod_{a=1}^t \prod_{j=1}^n  \diff W_{j,a} \right] \int_{-\infty}^{+\infty} \left[\prod_{a=1}^t \frac{\diff E_a}{2\pi} \right]\\
&\times \int_{-\infty}^{+\infty} \left[\prod_{a<b} \frac{\diff F_{ab} \diff Q_{ab}}{2\pi} \right]  e^{i  \sum_{a} E_a \left( \sum_j W_{j,a}^2 - n \right)}\\
&\times  e^{ i  \sum_{a<b}F_{ab}\left( \sum_j W_{j,a}W_{j,b} - n Q_{ab} \right) }\!  \int_{0}^{+\infty} \left[\prod_{a=1}^t \prod_{\mu=1}^p \frac{\diff \lambda^\mu_a \diff \bar{\lambda}^\mu_a}{(2\pi)^2} \right] \\
&\times \int_{-\infty}^{+\infty} \left[\prod_{a=1}^t \prod_{\mu=1}^p \diff x^\mu_a \diff \bar{x}^\mu_a \right]  e^{  i \sum_{a,\mu} x_a^\mu \left(\lambda^\mu_a - \frac{\sigma^\mu_a}{\sqrt{n}} \sum_{j} W_{j,a} \xi_j^\mu \right)}\\
&\times e^{ i \sum_{a,\mu} \bar{x}_a^\mu \left(\bar{\lambda}^\mu_a - \frac{\sigma^\mu_a}{\sqrt{n}} \sum_{j} W_{j,a} \bar{\xi}_j^\mu \right)}\,,
\end{aligned} 
\label{eq:replica_volume_replicated}
\end{equation}
where $1\le a,b\le t$ are replica indices (not to be confused with the indices running inside the multiplets, a notation we abandoned at the beginning of this section, when we specialized our calculation to doublets), $Q_{ab}$ is the replica matrix (with $ Q_{aa}=1 $) and $F_{ab}$ are the Lagrange multipliers enforcing the constraint
\begin{equation}
Q_{ab} = \frac{1}{n} \sum_{j=1}^n W_{j,a} W_{j,b} \,.
\label{eq:replica_matrix}
\end{equation}
Now we can perform the average over the input ensemble. With the same steps we used to get equation \eqref{eq:replica_average_annealed} (see Appendix~\ref{app:input}), we obtain, for the $x$ and $\lambda$ integrals,
\begin{multline}
\Biggl\{ \sum_{\{\sigma_a = \pm 1\}} \int_{0}^{+\infty} \left[\prod_{a=1}^t  \frac{ \Diff2 \bm{\lambda}_a}{(2\pi)^2} \right] \int_{-\infty}^{+\infty} \left[\prod_{a=1}^t  \Diff2 \bm{x}_a \right]\\
\times e^{-\frac{1}{2} \sum_{a,b} Q_{ab}  \bm{x}_a^T \mathcal{R} \bm{x}_b + i \sum_{a} \sigma_a \bm{x}^T_a \bm{\lambda}_a} \Biggr\}^p\,,
\label{eq:replica_bivariate}
\end{multline}
where we already inserted the replica matrix using \eqref{eq:replica_matrix} and we isolated the outputs $\sigma$ in the source term via the transformation $\bm{x} \to \sigma \bm{x}$. The remaining integral over the weights is Gaussian:
\begin{multline}
 \int  \left[\prod_{a,j} \diff W_{j,a} \right] e^{ i \sum_{a} E_a \sum_j W_{j,a}^2  + i \sum_{a<b} F_{ab} \sum_j W_{j,a}W_{j,b} }\\
= e^{-\frac{n}{2} \log \det\left(-i G \right)  + \frac{ n t}{2} \log(2\pi)}\,,
\end{multline}
where $G$ is the symmetric matrix with elements
\begin{equation}
G_{ab} = 2 E_a \delta_{ab} - (1-\delta_{ab})F_{ab}\,.
\end{equation}
The integral over the elements of $G$ is performed via a saddle-point: ignoring all the inessential factors,
\begin{multline}
\int_{-\infty}^{+\infty}\left[\prod_{a=1}^t \frac{\diff G_{aa}}{4\pi} \right]\int_{-\infty}^{+\infty} \left[\prod_{a<b} \frac{\diff G_{ab}}{2\pi} \right]\\
\times e^{-\frac{n}{2} \sum_{a,b} i G_{ab} Q_{ab}-\frac{n}{2} \log \det\left(-i G \right)}\\
 \sim e^{\frac{n t}{2}   +\frac{n}{2} \log \det(Q)}\,,
\end{multline}
where we used
\begin{equation}
\frac{\partial }{\partial G_{ab}} \left[\sum_{c,d} i G_{cd} Q_{cd}+ \log \det\left(-i G \right) \right] = i Q_{ab} + \left(G^{-1} \right)_{ba}\,.
\end{equation}
Finally, the resulting averaged replicated volume to be evaluated is
\begin{equation}
\begin{aligned}
&\mean{V^t} =  \int_{-\infty}^{+\infty} \left[\prod_{a<b} \diff Q_{ab} \right] e^{\frac{n t}{2}   +\frac{n}{2} \log \det(Q)}\\
&\times\Biggl\{ \sum_{\{\sigma_a = \pm 1\}} \int_{0}^{+\infty} \left[\prod_{a=1}^t  \frac{\Diff2 \bm{\lambda}_a }{(2\pi)^2} \right] \int_{-\infty}^{+\infty} \left[\prod_{a=1}^t  \Diff2 \bm{x}_a \right]\\
&\qquad\times e^{-\frac{1}{2} \sum_{a,b} Q_{ab} \bm{x}_a^T \mathcal{R}
\bm{x}_b + i \sum_{a} \sigma_a \bm{x}_a^T \bm{\lambda}_a} \Biggr\}^p\,.
\end{aligned} 
\label{eq:replica_volume_replicated_av}
\end{equation}
We cannot proceed further, in taking the limit $t\to 0$ as prescribed by the replica approach, without making an ansatz on the form of the replica matrix $Q_{ab}$.

\subsubsection{RS ansatz}
In the RS ansatz, the replica matrix has the form
\begin{equation}
Q_{ab} = (1-q)\delta_{ab} + q\,,\quad 0\le q\le 1\,,
\end{equation}
so that 
\begin{equation}
\log \det (Q) \underset{t\to 0}{\to} t \log(1-q) + \frac{t q}{1-q}\,.
\label{eq:replica_RS_measure}
\end{equation}
The quadratic form at the exponent of Eq.~\eqref{eq:replica_volume_replicated_av} reads
\begin{equation}
\begin{aligned}
&  \sum_{a,b}Q_{ab} \bm{x}_a^T \mathcal{R}\bm{x}_b \\
&= (1-q)\sum_{a} \bm{x}_a^T \mathcal{R}\bm{x}_a + q \left(\sum_{a}\bm{x}_a\right)^T \mathcal{R} \left(\sum_b  \bm{x}_b\right)\,.
\end{aligned}
\end{equation}
The last term can be linearized with a Hub\-bard-Stra\-to\-no\-vich transformation:
\begin{equation}
\begin{aligned}
&e^{- \frac{q}{2}  \left[\sum_{a}\bm{x}_a\right]^T \mathcal{R} \left[\sum_b  \bm{x}_b\right]} \\
&\qquad=\int_{-\infty}^{+\infty}\!  \frac{\Diff2 \bm{y}}{2\pi\sqrt{1-\rho^2}} \, e^{-\frac{1}{2} \bm{y}^T \mathcal{R}^{-1} \bm{y} + i\sqrt{q} \sum_a\bm{x}_a^T \bm{y}}\,.
\end{aligned}
\end{equation}
so that replica indices factorise, to get, after an integration over $\bm{x}$,
\begin{equation}
\begin{aligned}
&\Biggl\{\int_{-\infty}^{+\infty}\! \frac{\Diff2 \bm{y}}{2\pi\sqrt{1-\rho^2}} \, e^{-\frac{1}{2} \bm{y}^T \mathcal{R}^{-1}\bm{y}} \Biggl[\frac{2\pi}{(1-q) \sqrt{1-\rho^2}}  \sum_{\{\sigma=\pm 1\}}\\
&\times \int_{0}^{+\infty} \frac{\Diff2 \bm{\lambda}}{(2\pi)^2}   e^{-\frac{1}{2(1-q)} (\bm{\lambda} + \sigma\sqrt{q} \bm{y})^T \mathcal{R}^{-1} (\bm{\lambda} + \sigma\sqrt{q} \bm{y})} \Biggr]^t \Biggr\}^p\,.
\end{aligned}
\end{equation}
Defining $L_{RS}(\bm{y})$ the quantity in square brackets, the limit $t\to 0$ gives
\begin{equation}
\begin{aligned}
&p \log\Biggl\{\int_{-\infty}^{+\infty}\! \frac{\Diff2 \bm{y}}{2\pi\sqrt{1-\rho^2}} \, e^{-\frac{1}{2} \bm{y}^T \mathcal{R}^{-1}\bm{y}} \left[L_{RS}(\bm{y}) \right]^t \Biggr\} \\
&\to p t \int_{-\infty}^{+\infty}\! \frac{\Diff2 \bm{y}}{2\pi\sqrt{1-\rho^2}} \, e^{-\frac{1}{2} \bm{y}^T \mathcal{R}^{-1}\bm{y}} \log\left[L_{RS}(\bm{y}) \right]\,.
\end{aligned}
\end{equation}

Since we are looking for the critical value of $\alpha$ of the SAT-UNSAT transition of our CSP, we can just apply the standard argument by Gardner~\cite{Gardner:1987}: starting with a load below the critical value and increasing the number of patterns, 
the set of solutions in the space of weights
shrinks down to a single configuration at the transition (in the thermodynamic limit). This means that, approaching the critical point, the replicas of the vector $W$ must be more and more correlated and therefore $q \to 1$ at the transition.
In this limit, the factor $(1-q)^{-1}$ is large and the integrals in $L_{RS}(\bm{y})$ can be evaluated with a saddle point: we need to find the stationary points of the exponent in the integrands as a function of $\bm{\lambda}$. According to the position of the vector $\bm{y}$ on the plane, the saddle is in one of the three following spots: (i) inside the region of integration over $\bm{\lambda}$; (ii) at one of its boundaries; (iii) at the origin. We obtain:
\begin{equation}
\begin{aligned}
&\int_{0}^{+\infty} \frac{\Diff2 \bm{\lambda} }{2\pi(1-q) \sqrt{1-\rho^2}} \,  e^{-\frac{1}{2(1-q)} (\bm{\lambda} + \sigma\sqrt{q} \bm{y})^T\mathcal{R}^{-1} (\bm{\lambda} + \sigma \sqrt{q} \bm{y})}\\
&\sim \theta(-\sigma y) \theta(-\sigma\bar{y})  
+ \theta(\sigma y) \theta[\sigma (\rho y - \bar{y}) ] \frac{e^{-\frac{y^2}{2(1-q)}}}{y} \sqrt{\frac{1-q}{8\pi}} \\
&+ \theta[\sigma(\rho \bar{y} - y)] \theta(\sigma\bar{y})  \frac{e^{-\frac{\bar{y}^2}{2(1-q)}}}{\bar{y}} \sqrt{\frac{1-q}{8\pi}}
+ e^{-\frac{1}{2(1-q)} \bm{y}^T\mathcal{R}^{-1}\bm{y}}\\
&\times \theta[\sigma(y -  \rho \bar{y})] \theta[\sigma(\bar{y}- \rho y)]\frac{1}{2\pi} \frac{(1-q)(1-\rho^2)^{3/2}}{(\bar{y} -  \rho y) (y -  \rho \bar{y})} \,,
\end{aligned}
\end{equation}
with the theta functions selecting in turn one of the above cases. In the summation over $\sigma=\pm 1$, in each domain of $\bm{y}$ survives only the dominant addend in $(1-q)$: this is the finite term in the first and third quadrant, and the terms proportional to 
$\exp\{-y^2/[2(1-q)]\}$ or $\exp\{-\bar{y}^2/[2(1-q)]\}$ 
in the second and forth quadrant (the quadrants bisectors discriminating the larger). In the end, using the obvious symmetry between $y$ and $\bar{y}$ as integration variables and ignoring suppressed factors in $(1-q)$, we get
\begin{equation}
\begin{aligned}
&\int_{-\infty}^{+\infty}\! \frac{\Diff2 \bm{y}}{2\pi\sqrt{1-\rho^2}} \, e^{-\frac{1}{2} \bm{y}^T \mathcal{R}^{-1}\bm{y}} \log\left[L_{RS}(\bm{y}) \right]\\
&=\int_{0}^{+\infty}\! \frac{\diff y}{\pi \sqrt{1-\rho^2}}\, \frac{-y^2}{1-q}\int_{-\infty}^{- y}\diff \bar{y}\, e^{-\frac{1}{2} (y,\,\bar{y}) \mathcal{R}^{-1} \begin{psmallmatrix}
y\\
\bar{y}
\end{psmallmatrix}} \\
&=\frac{1}{4(1-q)} \left(\frac{2}{\pi}\sqrt{1-\rho^2} - \frac{4}{\pi} \arctan \frac{\sqrt{1-\rho}}{\sqrt{1+\rho}} \right)\,.
\end{aligned}
\label{eq:replica_RS_L}
\end{equation}
Selecting only the most divergent terms in $(1-q)$ from \eqref{eq:replica_RS_measure} and \eqref{eq:replica_RS_L}, we have all the ingredients to evaluate the replica limit of $(\mean{V^t}-1)/t$ for $t\to 0$. The result is zero when the load $\alpha$ assumes the critical value
\begin{equation}
\alpha_*^{\text{RS}} (\rho) = \frac{\pi}{2 \arctan \sqrt{(1-\rho)/(1+\rho)} -\sqrt{1-\rho^2}}\,.
\label{eq:replica_alphaRS}
\end{equation}

The result is reported in Fig.~\ref{fig:replica_alpha}: the RS curve presents the expected limits~\eqref{eq:alpha_limits}, but again we do not observe quantitative agreement with the combinatorial curve. 
We are therefore led to conjecture that we need at least one step of replica symmetry breaking (RSB). We work out the derivation of $\alpha_{\ast}$ within the $1$RSB ansatz in the next section.

\subsubsection{1RSB ansatz}
In the 1RSB ansatz the replica matrix has the form
\begin{equation}
Q_{ab} = (1-q_1) \delta_{ab} +(q_1 -q_0 ) \varepsilon_{ab} + q_0\,,
\end{equation}
where $\varepsilon_{ab}=1$ if $a$, $b$ belongs to a diagonal block $m\times m$, 0 otherwise, so that
\begin{equation}
\begin{aligned}
&\log \det (Q) \to t \biggl\{\frac{m-1}{m}\log(1-q_1) \\
&+\frac{1}{m}\log[1-q_1+m(q_1-q_0)] + \frac{q_0}{1-q_1+m(q_1-q_0)} \biggr\}\,.
\end{aligned}
\end{equation}
From \eqref{eq:replica_volume_replicated_av}, we get
\begin{equation}
\begin{aligned}
&  \sum_{a,b} Q_{ab}\bm{x}^T_a \mathcal{R}\bm{x}_b 
= (1-q_1)\sum_{a} \bm{x}^T_a \mathcal{R}\bm{x}_a\\
&\quad + (q_1 - q_0) \sum_{B=0}^{t/m-1} \left(\sum_{a=1}^m \bm{x}_{mB+a}\right)^T \mathcal{R}\left(\sum_{b=1}^{m} \bm{x}_{mB+b}\right)\\
&\quad+ q_0  \left(\sum_{a} \bm{x}_{a}\right)^T \mathcal{R} \left(\sum_{b}  \bm{x}_{b}\right)\,,
\end{aligned}
\end{equation}
where $B$ is a block index. We now need $2(t/m+1)$ auxiliary Hubbard-Stratonovich variables to linearize the sums over replica indices:
to get, after the usual factorisations and the integration over $\bm{x}$,
\begin{equation}
\begin{aligned}
&\Biggl\{\int \frac{\Diff2{\bm{y}}\,e^{-\frac{1}{2} \bm{y}^T  \mathcal{R}^{-1} \bm{y}}}{2\pi \sqrt{1-\rho^2}} \Biggl[\int\frac{\Diff2{\bm{z}}\,e^{-\frac{1}{2} \bm{z}^T  \mathcal{R}^{-1} \bm{z}}}{2\pi \sqrt{1-\rho^2}}\\
&\times \Biggl(\sum_{\{\sigma=\pm 1 \}} \int_0^{+\infty}\frac{\Diff2 \bm{\lambda}}{2\pi (1-q_1)\sqrt{1-\rho^2}} \\
&\times  e^{-\frac{ \left[\sigma\left( \sqrt{q_1-q_0} \bm{z}  +  \sqrt{q_0} \bm{y}\right) +  \bm{\lambda} \right]^T \mathcal{R}^{-1}\left[ \sigma\left(\sqrt{q_1-q_0} \bm{z}  +  \sqrt{q_0} \bm{y}\right) + \bm{\lambda} \right]}{2(1-q_1)}} \Biggr)^m \Biggr]^{\frac{t}{m}} \Biggr\}^p\!.
\end{aligned}
\end{equation}
Defining $L_{1RSB}(\bm{y})$ the argument of the square brackets, we know that the logarithm of the above quantity for $t\to 0$ gives
\begin{equation}
\begin{aligned}
&\frac{pt}{m}\int \frac{\Diff2{\bm{y}}\,e^{-\frac{1}{2} \bm{y}^T \mathcal{R}^{-1} \bm{y}}}{2\pi \sqrt{1-\rho^2}}  \,\log\left[L_{1RSB}(\bm{y}) \right]\,.
\end{aligned}
\end{equation}
To simplify $L_{1RSB}(\bm{y})$ and to get an expression similar to the one we studied before, we can shift the $\bm{z}$ variables to
\begin{equation}
\bm{z} \to \bm{z} - \frac{\sqrt{q_0}}{\sqrt{q_1-q_0}}\bm{y}\,,
\end{equation}
obtaining
\begin{equation}
\begin{aligned}
&L_{1RSB}(\bm{y}) = \int\frac{\Diff2{\bm{z}}\,e^{-\frac{1}{2} \left(\bm{z} - \frac{\sqrt{q_0}}{\sqrt{q_1-q_0}}\bm{y}\right)^T \mathcal{R}^{-1} \left(\bm{z} - \frac{\sqrt{q_0}}{\sqrt{q_1-q_0}}\bm{y}\right)}}{2\pi \sqrt{1-\rho^2}} \\
&\times \Biggl(\sum_{\{\sigma=\pm 1 \}} \int_0^{+\infty} \frac{\Diff2 \bm{\lambda} \, e^{-\frac{ \left[\sigma\sqrt{q_1-q_0} \bm{z}   +  \bm{\lambda} \right]^T \mathcal{R}^{-1}\left[ \sigma\sqrt{q_1-q_0} \bm{z}  + \bm{\lambda} \right]}{2(1-q_1)}}}{2\pi(1-q_1)\sqrt{1-\rho^2}} \Biggr)^m \!\!.
\end{aligned} 
\end{equation}
In order to find the critical load, we investigate the behaviour of the 1RSB parameters close to the transition: it turns out that $q_1$ has to be sent to one (in analogy with the RS case) and $m$ to zero~\cite{Malatesta_2019} as 
\begin{equation}
q_1 \to 1\,,\qquad m \to (1-q_1)w\,,
\label{eq:replica_1RSBlimit}
\end{equation}
with $w$ a finite parameter. In this limit we can evaluate the integral over $\bm{\lambda}$ with a saddle point. We get
\begin{equation}
\theta(z) \theta(\bar{z}) + \theta(-z) \theta(-\bar{z}) + 4 \theta(z) \theta(-\bar{z}-z) e^{- \frac{w(1-q_0)z^2}{2}}\,.
\end{equation}
Analytical computations are rather cumbersome after this point. However, the result simplifies a lot if we take $q_0 = 0$. Then the integral over $\bm{y}$ decouples and simply gives 1, while the one over $\bm{z}$ breaks into the regions
\begin{equation}
\int_{0}^{+\infty}\frac{ \Diff2  \bm{z}}{\pi \sqrt{1-\rho^2}}\,e^{-\frac{1}{2}\bm{z}^T\mathcal{R}^{-1}\bm{z}} = \frac{1}{2} + \frac{1}{\pi} \arcsin (\rho) 
\end{equation}
and
\begin{multline}
 \int_{0}^{+\infty} \frac{2 \diff z}{\pi \sqrt{1-\rho^2}} \int_{-\infty}^{-z} \diff \bar{z} \,e^{-\frac{1}{2}\bm{z}^T\mathcal{R}^{-1}\bm{z}- \frac{w z^2}{2}} \\
 = \frac{2 \arctan\left(\sqrt{(1+w)\frac{ 1 - \rho}{1+\rho}}\right)}{\pi  \sqrt{1+w}} \,.
\end{multline}
In the end, we find
\begin{multline}
\alpha_*^{\text{1RSB}}(\rho;q_0=0,w) \\
=  \frac{-\log[1+w] }{ 2 \log \left[\frac{1}{2} + \frac{1}{\pi} \arcsin (\rho) +\frac{2 \arctan\left(\sqrt{(1+w)\frac{ 1 - \rho}{1+\rho}}\right)}{\pi  \sqrt{1+w}} \right]}\,.
\label{eq:replica_alpha1RSB}
\end{multline}
We stress that this last result is not the optimal 1RSB solution: in principle we should consider the full expression of $\alpha_*^{\text{1RSB}}(\rho;q_0,w)$ and optimize upon the remaining parameters $q_0$ and $w$. However, this is beyond the scope of this section: here, we simply verify that the functional form $\alpha_*^{\text{1RSB}}(\rho;q_0=0,w)$ allows to fit nicely the combinatorial result, by adjusting the parameter $w$ (see Fig.~\ref{fig:replica_alpha}). This simple observation strongly supports our conjecture that this SAT-UNSAT transition exhibits at least one step of RSB, but it does not rule out a full-RSB scenario.

\subsection{Margin learning}
\label{sec:spheres}

Replica theory turns out to be essential to explore the role of data structure whenever alternative, \emph{ad hoc} methods (such as the combinatorial one) are not available. Here we apply it to identify the SAT-UNSAT transition occurring in 
margin learning.
The synaptic volume relevant to this case is
\begin{equation}
\begin{aligned}
V_{\kappa} {}={}& \sum_{\{\sigma^\mu = \pm1\}} \int \left[ \prod_{j=1}^n \diff W_j \right]\delta\! \left(\sum_{j=1}^n W_j^2 -n \right) \\
&\times \prod_{\mu=1}^{p} \theta\!\left(\frac{\sigma^\mu}{\sqrt{n}} \sum_{j=1}^n W_j \xi_j^\mu - \kappa \right)\,,
\end{aligned}
\end{equation}
where $\kappa$ is the margin. Note again that here, as in the case of Eq.~\eqref{eq:replica_volume}, the outputs $\sigma^\mu$ are dynamical variables, at variance with the usual Gardner's volume. We skip the details on the annealed and quenched calculations, which are in spirit very similar to those of the previous sections. Nonetheless, it is worth to point out that the tricky multivariate integrals in the auxiliary variable, are now replaced by Gaussian integrals, with the margin $\kappa$ appearing as an integration limit. The annealed approximation leads to 
\begin{equation}
\alpha_*^{\text{A}}(\kappa) = - \frac{1+ \log(2\pi)}{2 \log[2\erfc(\kappa)]}\,.
\label{eq:margin_alphaA}
\end{equation}
In the quenched calculation, the RS ansatz is again implemented by requiring $q\to 1$; 
one obtains the critical threshold
\begin{equation}
\alpha_*^{\text{RS}}(\kappa) = \frac{1}{2} \left[\int_{0}^{\kappa}\!\dgau y\,(\kappa - y)^2  \right]^{-1}\,,
\label{eq:margin_alphaRS}
\end{equation}
where $\dgau y$ is the Gaussian measure. Note the difference with Gardner's result~\cite{Gardner:1987} for the storage capacity,
\begin{equation}
\alpha_c(\kappa) = \left[\int_{-\kappa}^{+\infty} \!\dgau y\left(\kappa + y\right)^2 \right]^{-1}\,.
\label{eq:margin_alphaC}
\end{equation}
The one-step RSB ansatz again depends on the parameters $q_0$ and $w$, which should be investigated numerically. However, in the special case $q_0=0$ we find the simpler expression
\begin{multline}
\alpha_*^{\text{1RSB}}(\kappa;q_0=0,w)\\
 = \frac{ - \log[1+w] }{\displaystyle 2 \log \left\{2\left[\erfc(\kappa) +\int_{0}^{\kappa} \dgau z\, e^{-w\frac{\left( z  - \kappa\right)^2}{2} }\right]\right\}}\,.
\end{multline}

These results essentially share the same features of those for the simplexes computed above: in particular, at variance with the usual storage capacity~\eqref{eq:margin_alphaC}, $\alpha_*$ computed in all the different approximation schemes diverges in the limit $\kappa\to 0^+$, when the problem reduces to a standard classification of points 
(or equivalently, in the object manifold description, when the radius of the spheres shrinks to zero). 
Even in absence of
a closed expression for the VC entropy of margin classification, 
the existence of the phase transition at a finite load is a clear indication of its non-monotonicity.


\section{Discussion}

Understanding how data specificities impact the performance of machine learning models and algorithms 
can be considered one of the major challenges for contemporary statistical physics. 
Here we have shown how to deal with data structure, 
as it is being established in physics,
within the framework of the statistical theory of learning.
The presence of input-output correlations in a dataset suggests
constraints to be applied to the hypothesis class under consideration.
As a result, the corresponding VC entropy, deeply connected to the generalization capabilities of the model,
is considerably lower than in the unstructured case.

For simple models of data structure we have observed two striking phenomena
that take place above the VC dimension.
First, the VC entropy becomes nonmonotonic.
This is a strong indication that the rigorous bounds in SLT may be substantially improved
by taking data structure into account.
Second, a novel transition appears beyond the well-known storage capacity, at
the onset of unsatisfiability for a data-related constraint satisfaction problem.
When available, a combinatorial theory \emph{\`a la Cover} allows one to compute the VC entropy of a finite-size system,
and to reveal explicitly its nonmonotonic behavior.
However, this is not always feasible, such as for spherical object manifolds and margin learning.
In these cases, we showed how the phase transition can be probed with the standard tools
of statistical physics, thus allowing an indirect quantification of the data-dependent behavior.

The new satisfiability transition is due to a competition between the combinatorial expansion,
with sample size,
of the space of possible functions and the reduction due to the constraints \cite{companion1}.
We believe, as this observation suggests, that the emergence of the data-driven transition,
as well as the nonmonotonic VC entropy it entails, is not specific to the two models of data that we have studied here,
but is more generally present whenever the constraints imposed on the hypothesis class
by data structure are strong enough.
On a more quantitative level, notice that the upper and lower bounds obtained for $\alpha_*$
in Sec.~\ref{section:generic_k} are very close to one another.
The bounds are independent of the particular choice of simplexes, i.e., they do not depend
on $k$ or on $\{\rho_{ab}\}$.
This is a clue pointing to the robustness of the phenomenology for disparate data structures.
We remark that the combinatoric analysis was done at leading order in $\alpha$;
thus, it remains to assess how much the bounds are affected by perturbative corrections.

An ambitious and pressing goal concerns the generalization of our results to other architectures,
notably deep neural networks,
in the same spirit of what was achieved in SLT regarding the VC dimension.

\begin{acknowledgments}
The authors would like to thank Enrico Malatesta and Marco Cosentino Lagomarsino for useful discussions and suggestions.
\end{acknowledgments}

\appendix

\section{Bounds on \texorpdfstring{$\theta_{1}^{k}$}{theta(1,k)}}
\label{app:bounds_theta}
In this Appendix we report the details of the calculation of the bounds on $\theta_1^k$ given in Section~\ref{section:generic_k}.
To briefly recall the definitions, we have 
\begin{equation}
    \begin{split}
        \theta_0^k &= \prod_{m=2}^k \psi_m \\
        \theta_1^k &= \left( 2-k + \sum_{m=2}^k \frac{1}{\psi_m} \right) \prod_{m=2}^k \psi_m   
        \, 
    \end{split}
\end{equation}
for some $0 \leq \psi_m \leq 1$, $\forall m \geq 2$.
We want to compute the infimum and the supremum of $\theta_1^k$ at fixed $\theta_0^k$, as a function of the $\psi$ variables.
First of all, let us simplify the notation. 
Define:
\begin{equation}
    \begin{split}
        x_m := \psi_{m+1} &\, ,\quad \forall m \geq 1\\
        f_{(k)}(x_1 \dots x_k) := \theta_1^{k+1}(\psi_2\dots\psi_{k+1}) &\, ,\quad \forall k \geq 1 \, .
    \end{split}
\end{equation}
Explicitly:
\begin{equation}
    \begin{split}
        f_{(k)}(\vec{x}) = \left( 1-k+\sum_{m=1}^k \frac{1}{x_m} \right) \prod_{m=1}^k x_m \, ,
    \end{split}
\end{equation}
where $\vec{x} = (x_1\dots x_k)$.

Our problem is to optimize (i.e., to find the infimum and the supremum) $f_{(k)}(\vec{x})$ in the hypercube $\vec{x}\in[0,1]^k$, subject to the constraint
\begin{equation}
    \begin{split}
        \prod_{m=1}^k x_m = t \in [0,1] \, .
    \end{split}
\end{equation}

We will prove by induction that 
\begin{equation}
    \begin{split}
        \sup_{\vec{x} \in [0,1]^k } f_{(k)}(\vec{x}) &= 1 - \delta_{t,0} \, ,\quad \forall k\geq 1 \\
        \inf_{\vec{x} \in [0,1]^k } f_{(k)}(\vec{x}) &= \phi(k,t) \, ,\quad \forall k\geq 1
            \, 
    \end{split}
\end{equation}
where $\phi(k,t) = (1-k) t + k t^{1-\frac{1}{k}}$.
Notice that $\phi(k,t)$ is a monotone decreasing function of $k$, and it is always less then $1$.

The case $t=0$ is special, as the constraint restricts the domain to the origin and $f_{(k)}$ is null; in the following, suppose that $t>0$.

If $k=1$, the constraint implies that $x_1=t$, so that $f_{(1)}(x_1) = f_{(1)}(t) = 1$.
The fact that $\phi(1,t) = 1$ proves that the proposed bounds are indeed true.

If $k>1$, we first look for critical points inside $[0,1]^k$ using Lagrange's theorem for constrained optimization; then, we optimize our function on the boundary of $[0,1]^k$ to look for non-critical extrema.
\begin{itemize}
    \item Inside the domain, Lagrange's theorem gives that $\vec{x}_* = (t^\frac{1}{k}\dots t^\frac{1}{k})$ is the only critical point, and $f_{(k)}(\vec{x}_*) = \phi(k,t)$;
    \item on the boundary, we have that at least one of the $x$ variables (without loss of generality, let us take $x_k$ to be this boundary variable) must be either $0$ or $1$; the former is not compatible with the constraint as $t>0$, so $x_k=1$.
        But $f_{(k)}(x_1\dots x_{k-1},1) = f_{(k-1)}(x_1 \dots x_{k-1})$, and $t = \prod_{m=1}^k x_m = \prod_{m=1}^{k-1}x_m$, so that the constrained optimization of $f_{(k)}(\vec{x})$ on the boundary of the domain is equivalent to the constrained optimization of $f_{(k-1)}(\vec{x})$ on the full domain $[0,1]^{k-1}$.
\end{itemize}
Thus, the candidates for the infimum and the supremum of $f_{(k)}(\vec{x})$ are given by $\phi(k)$ (inside the domain, by Lagrange's theorem) and $1,\phi(k-1,t)$ (on the boundary of the domain, by induction hypothesis).
The properties of $\phi$ imply that $1$ is the supremum and $\phi(k,t)$ is the infimum of $f_{(k)}(\vec{x})$.

Finally, again by induction, we see that the supremum is realized on the point $(t,1,\dots)$ and by all the distinct permutations of its coordinates, and that the infimum is realized by $(t^\frac{1}{k}\dots t^\frac{1}{k})$.

\section{Averaging over the input distribution}
\label{app:input}
In this Appendix we report the details of the calculation of the averages over the input ensemble, performed in Sec.~\ref{replicares}. From Eq.~\eqref{eq:replica_inputs}, specialized for $k=2$, we observe that at fixed overlap $\rho$, given  $c$, $d$ $\in \mathbb{N}$ the numbers of concordant and discordant signs of the components of the pair for each $\mu$, then $c - d = \rho n$, $c + d = n$, so
\begin{equation}
c = (1+\rho)n/2\,,\qquad d = (1-\rho)n/2\,.
\label{eq:replica_concordance}
\end{equation}
For each $\mu$, we can freely choose in $2^n$ different ways the components of $\xi^\mu$, but then for $\bar{\xi}^\mu$ we must take $c$ components with the same sign of their counterparts and $d$ with the opposite. We can do that in $\binom{n}{c}$ different ways, so the normalization factor is
\begin{equation}
\nu = 2^{pn}\binom{n}{\frac{(1+\rho)n}{2}}^p.
\end{equation}
However, the order of the components of the vectors $\xi$, $\bar{\xi}$ is completely irrelevant, because they appear only in scalar products, among themselves (in the overlap constraint) and with the same vector $W$, whose components again we are free to relabel. This means that we can choose as a representative of the vector $\bar{\xi}$, for example, the one with the concordant components at the beginning. We can write the ensemble measure as
\begin{equation}
\begin{aligned}
\diff P_\rho (\Xi) = \prod_{\mu=1}^p  {}&{}\left[\prod_{j=1}^c\diff P(\xi^\mu_j) \delta(\xi^\mu_j - \bar{\xi}^\mu_j) \diff \bar{\xi}^\mu_j\right]\\
\times{}&{}\left[\prod_{j=c+1}^n\diff P(\xi^\mu_j) \delta(\xi^\mu_j + \bar{\xi}^\mu_j) \diff \bar{\xi}^\mu_j\right]\,,
\label{eq:replica_inputs_k2}
\end{aligned}
\end{equation}
where
\begin{equation}
\diff P(\xi^\mu_j) = \frac{1}{2} \left[\delta(\xi^\mu_{j} - 1) + \delta(\xi^\mu_{j} +1)\right]   \diff \xi_{j}^\mu \,.
\end{equation}
Note that with the choice of a representative we are explicitly breaking the invariance of the original expression under permutation (relabeling) of the indices $j$, a symmetry we will reintroduce by hand in the following calculation. 

We can now perform the averages of the volumes~\eqref{eq:replica_volume_k2} and~\eqref{eq:replica_volume_replicated}. We report only the annealed calculation, the quenched one being a straightforward variation. Isolating the only part depending on the inputs in the integrand of Eq.~\eqref{eq:replica_volume_k2}, we find
\begin{equation}
\begin{aligned}
&\int \diff P_\rho(\Xi)\,e^{ - i \sum_{\mu}\sigma^\mu x^\mu \sum_{j}  \frac{\xi_j^\mu W_{j} }{\sqrt{n}}  - i \sum_{\mu} \sigma^\mu \bar{x}^\mu \sum_{j} \frac{\bar{\xi}_j^\mu W_{j} }{\sqrt{n}}  } \\
&= \prod_{\mu=1}^p \prod_{j=1}^c  \cos \left[\frac{1}{\sqrt{n}} \left(x^\mu+ \bar{x}^\mu  \right)\sigma^\mu  W_{j}   \right]\\
&\phantom{={}}\times\prod_{j=c+1}^n  \cos \left[\frac{1}{\sqrt{n}} \left(x^\mu- \bar{x}^\mu  \right)\sigma^\mu  W_{j}   \right]\\
&\approx \prod_{\mu=1}^p e^ {-\frac{1}{2} (x^\mu)^2 \sum_{j=1}^n \frac{W_{j}^2}{n} -\frac{1}{2} (\bar{x}^\mu)^2  \sum_{j=1}^n \frac{W_{j}^2}{n} }\\
&\phantom{={}}\times e^{-x^\mu \bar{x}^\mu \left(\sum_{j=1}^c  -\sum_{j=c+1}^n \right)\frac{W_{j}^2}{n} }\,,
\end{aligned}
\end{equation}
where, in the final step, a large $n$ expansion is performed. The exponent of the last term, consisting in a sum over $j$ that does not extend over all the $n$ components, cannot be readily solved using the spherical constraint, but we can write it as
\begin{equation}
\biggl(\sum_{j=1}^c  - \sum_{j=c+1}^n\biggr) \frac{W_{j}^2}{n} = \biggl(2\sum_{j=1}^c  - \sum_{j=1}^n\biggr) \frac{W_{j}^2}{n}\,.
\end{equation}
Now, only the first sum is not invariant under permutations of the components. However, since the starting point was symmetric, we can also multiply this expression by similar ones obtained with other choices of the vector $\bar{\xi}^\mu$, and then take the corresponding root of the result, obtaining an equivalent formula. The trick to restore a complete sum over the $n$ components, is to multiply by all the $c$-permutations of $n$, and then take the $n!/(n-c)!$-th root of the result. The only non-trivial term at the exponent during this procedure is indeed the partial sum, which reads:
\begin{equation}
\begin{aligned}
\frac{(n-c)!}{c!} \sum_{j=1}^c\sum_{\substack{\pi_1\neq \pi_2 \neq \cdots \neq \pi_c\\\forall i,\,1\le\pi_i\le n}}  \frac{W_{\pi_j}^2}{n} 
= \frac{c}{n}\sum_{i=1}^n \frac{W_{i}^2}{n}\,.
\end{aligned}
\end{equation}
Now the spherical constraint can be invoked on all terms. Using  $\left(2c/n-1\right) = \rho$, and factorising the $p$ integrals over the auxiliary variables $x$ and $\lambda$, we obtain Eq.~\eqref{eq:replica_average_annealed}.

\bibliographystyle{apsrev4-1}
\bibliography{biblio}

\begin{thebibliography}{47}%
\makeatletter
\providecommand \@ifxundefined [1]{%
 \@ifx{#1\undefined}
}%
\providecommand \@ifnum [1]{%
 \ifnum #1\expandafter \@firstoftwo
 \else \expandafter \@secondoftwo
 \fi
}%
\providecommand \@ifx [1]{%
 \ifx #1\expandafter \@firstoftwo
 \else \expandafter \@secondoftwo
 \fi
}%
\providecommand \natexlab [1]{#1}%
\providecommand \enquote  [1]{``#1''}%
\providecommand \bibnamefont  [1]{#1}%
\providecommand \bibfnamefont [1]{#1}%
\providecommand \citenamefont [1]{#1}%
\providecommand \href@noop [0]{\@secondoftwo}%
\providecommand \href [0]{\begingroup \@sanitize@url \@href}%
\providecommand \@href[1]{\@@startlink{#1}\@@href}%
\providecommand \@@href[1]{\endgroup#1\@@endlink}%
\providecommand \@sanitize@url [0]{\catcode `\\12\catcode `\$12\catcode
  `\&12\catcode `\#12\catcode `\^12\catcode `\_12\catcode `\%12\relax}%
\providecommand \@@startlink[1]{}%
\providecommand \@@endlink[0]{}%
\providecommand \url  [0]{\begingroup\@sanitize@url \@url }%
\providecommand \@url [1]{\endgroup\@href {#1}{\urlprefix }}%
\providecommand \urlprefix  [0]{URL }%
\providecommand \Eprint [0]{\href }%
\providecommand \doibase [0]{http://dx.doi.org/}%
\providecommand \selectlanguage [0]{\@gobble}%
\providecommand \bibinfo  [0]{\@secondoftwo}%
\providecommand \bibfield  [0]{\@secondoftwo}%
\providecommand \translation [1]{[#1]}%
\providecommand \BibitemOpen [0]{}%
\providecommand \bibitemStop [0]{}%
\providecommand \bibitemNoStop [0]{.\EOS\space}%
\providecommand \EOS [0]{\spacefactor3000\relax}%
\providecommand \BibitemShut  [1]{\csname bibitem#1\endcsname}%
\let\auto@bib@innerbib\@empty
\bibitem [{\citenamefont {Amit}\ \emph
  {et~al.}(1985{\natexlab{a}})\citenamefont {Amit}, \citenamefont {Gutfreund},\
  and\ \citenamefont {Sompolinsky}}]{PhysRevA.32.1007}%
  \BibitemOpen
  \bibfield  {author} {\bibinfo {author} {\bibfnamefont {D.~J.}\ \bibnamefont
  {Amit}}, \bibinfo {author} {\bibfnamefont {H.}~\bibnamefont {Gutfreund}}, \
  and\ \bibinfo {author} {\bibfnamefont {H.}~\bibnamefont {Sompolinsky}},\
  }\href {\doibase 10.1103/PhysRevA.32.1007} {\bibfield  {journal} {\bibinfo
  {journal} {Phys. Rev. A}\ }\textbf {\bibinfo {volume} {32}},\ \bibinfo
  {pages} {1007} (\bibinfo {year} {1985}{\natexlab{a}})}\BibitemShut {NoStop}%
\bibitem [{\citenamefont {Amit}\ \emph
  {et~al.}(1985{\natexlab{b}})\citenamefont {Amit}, \citenamefont {Gutfreund},\
  and\ \citenamefont {Sompolinsky}}]{PhysRevLett.55.1530}%
  \BibitemOpen
  \bibfield  {author} {\bibinfo {author} {\bibfnamefont {D.~J.}\ \bibnamefont
  {Amit}}, \bibinfo {author} {\bibfnamefont {H.}~\bibnamefont {Gutfreund}}, \
  and\ \bibinfo {author} {\bibfnamefont {H.}~\bibnamefont {Sompolinsky}},\
  }\href {\doibase 10.1103/PhysRevLett.55.1530} {\bibfield  {journal} {\bibinfo
   {journal} {Phys. Rev. Lett.}\ }\textbf {\bibinfo {volume} {55}},\ \bibinfo
  {pages} {1530} (\bibinfo {year} {1985}{\natexlab{b}})}\BibitemShut {NoStop}%
\bibitem [{\citenamefont {Gardner}(1987)}]{Gardner:1987}%
  \BibitemOpen
  \bibfield  {author} {\bibinfo {author} {\bibfnamefont {E.}~\bibnamefont
  {Gardner}},\ }\href {\doibase 10.1209/0295-5075/4/4/016} {\bibfield
  {journal} {\bibinfo  {journal} {Europhysics Letters ({EPL})}\ }\textbf
  {\bibinfo {volume} {4}},\ \bibinfo {pages} {481} (\bibinfo {year}
  {1987})}\BibitemShut {NoStop}%
\bibitem [{\citenamefont {Gardner}\ and\ \citenamefont
  {Derrida}(1988)}]{GardnerDerrida:1988}%
  \BibitemOpen
  \bibfield  {author} {\bibinfo {author} {\bibfnamefont {E.}~\bibnamefont
  {Gardner}}\ and\ \bibinfo {author} {\bibfnamefont {B.}~\bibnamefont
  {Derrida}},\ }\href {http://stacks.iop.org/0305-4470/21/i=1/a=031} {\bibfield
   {journal} {\bibinfo  {journal} {Journal of Physics A: Mathematical and
  General}\ }\textbf {\bibinfo {volume} {21}},\ \bibinfo {pages} {271}
  (\bibinfo {year} {1988})}\BibitemShut {NoStop}%
\bibitem [{\citenamefont {Chung}\ \emph {et~al.}(2018)\citenamefont {Chung},
  \citenamefont {Lee},\ and\ \citenamefont
  {Sompolinsky}}]{ChungLeeSompolinsky:2018}%
  \BibitemOpen
  \bibfield  {author} {\bibinfo {author} {\bibfnamefont {S.}~\bibnamefont
  {Chung}}, \bibinfo {author} {\bibfnamefont {D.~D.}\ \bibnamefont {Lee}}, \
  and\ \bibinfo {author} {\bibfnamefont {H.}~\bibnamefont {Sompolinsky}},\
  }\href {\doibase 10.1103/PhysRevX.8.031003} {\bibfield  {journal} {\bibinfo
  {journal} {Phys. Rev. X}\ }\textbf {\bibinfo {volume} {8}},\ \bibinfo {pages}
  {031003} (\bibinfo {year} {2018})}\BibitemShut {NoStop}%
\bibitem [{\citenamefont {Chung}\ \emph {et~al.}(2016)\citenamefont {Chung},
  \citenamefont {Lee},\ and\ \citenamefont
  {Sompolinsky}}]{ChungLeeSompolinsky:2016}%
  \BibitemOpen
  \bibfield  {author} {\bibinfo {author} {\bibfnamefont {S.}~\bibnamefont
  {Chung}}, \bibinfo {author} {\bibfnamefont {D.~D.}\ \bibnamefont {Lee}}, \
  and\ \bibinfo {author} {\bibfnamefont {H.}~\bibnamefont {Sompolinsky}},\
  }\href {\doibase 10.1103/PhysRevE.93.060301} {\bibfield  {journal} {\bibinfo
  {journal} {Phys. Rev. E}\ }\textbf {\bibinfo {volume} {93}},\ \bibinfo
  {pages} {060301} (\bibinfo {year} {2016})}\BibitemShut {NoStop}%
\bibitem [{\citenamefont {Cohen}\ \emph {et~al.}(2020)\citenamefont {Cohen},
  \citenamefont {Chung}, \citenamefont {Lee},\ and\ \citenamefont
  {Sompolinsky}}]{Cohen2020}%
  \BibitemOpen
  \bibfield  {author} {\bibinfo {author} {\bibfnamefont {U.}~\bibnamefont
  {Cohen}}, \bibinfo {author} {\bibfnamefont {S.}~\bibnamefont {Chung}},
  \bibinfo {author} {\bibfnamefont {D.~D.}\ \bibnamefont {Lee}}, \ and\
  \bibinfo {author} {\bibfnamefont {H.}~\bibnamefont {Sompolinsky}},\ }\href
  {\doibase 10.1038/s41467-020-14578-5} {\bibfield  {journal} {\bibinfo
  {journal} {Nature Communications}\ }\textbf {\bibinfo {volume} {11}},\
  \bibinfo {pages} {746} (\bibinfo {year} {2020})}\BibitemShut {NoStop}%
\bibitem [{\citenamefont {M\'ezard}(2017)}]{Mezard:2017}%
  \BibitemOpen
  \bibfield  {author} {\bibinfo {author} {\bibfnamefont {M.}~\bibnamefont
  {M\'ezard}},\ }\href {\doibase 10.1103/PhysRevE.95.022117} {\bibfield
  {journal} {\bibinfo  {journal} {Phys. Rev. E}\ }\textbf {\bibinfo {volume}
  {95}},\ \bibinfo {pages} {022117} (\bibinfo {year} {2017})}\BibitemShut
  {NoStop}%
\bibitem [{\citenamefont {Mazzolini}\ \emph
  {et~al.}(2018{\natexlab{a}})\citenamefont {Mazzolini}, \citenamefont
  {Gherardi}, \citenamefont {Caselle}, \citenamefont {Cosentino~Lagomarsino},\
  and\ \citenamefont {Osella}}]{Mazzolini:2018:PRX}%
  \BibitemOpen
  \bibfield  {author} {\bibinfo {author} {\bibfnamefont {A.}~\bibnamefont
  {Mazzolini}}, \bibinfo {author} {\bibfnamefont {M.}~\bibnamefont {Gherardi}},
  \bibinfo {author} {\bibfnamefont {M.}~\bibnamefont {Caselle}}, \bibinfo
  {author} {\bibfnamefont {M.}~\bibnamefont {Cosentino~Lagomarsino}}, \ and\
  \bibinfo {author} {\bibfnamefont {M.}~\bibnamefont {Osella}},\ }\href
  {\doibase 10.1103/PhysRevX.8.021023} {\bibfield  {journal} {\bibinfo
  {journal} {Phys. Rev. X}\ }\textbf {\bibinfo {volume} {8}},\ \bibinfo {pages}
  {021023} (\bibinfo {year} {2018}{\natexlab{a}})}\BibitemShut {NoStop}%
\bibitem [{\citenamefont {Mazzolini}\ \emph
  {et~al.}(2018{\natexlab{b}})\citenamefont {Mazzolini}, \citenamefont
  {Grilli}, \citenamefont {De~Lazzari}, \citenamefont {Osella}, \citenamefont
  {Lagomarsino},\ and\ \citenamefont {Gherardi}}]{Mazzolini:2018:PRE}%
  \BibitemOpen
  \bibfield  {author} {\bibinfo {author} {\bibfnamefont {A.}~\bibnamefont
  {Mazzolini}}, \bibinfo {author} {\bibfnamefont {J.}~\bibnamefont {Grilli}},
  \bibinfo {author} {\bibfnamefont {E.}~\bibnamefont {De~Lazzari}}, \bibinfo
  {author} {\bibfnamefont {M.}~\bibnamefont {Osella}}, \bibinfo {author}
  {\bibfnamefont {M.~C.}\ \bibnamefont {Lagomarsino}}, \ and\ \bibinfo {author}
  {\bibfnamefont {M.}~\bibnamefont {Gherardi}},\ }\href {\doibase
  10.1103/PhysRevE.98.012315} {\bibfield  {journal} {\bibinfo  {journal} {Phys.
  Rev. E}\ }\textbf {\bibinfo {volume} {98}},\ \bibinfo {pages} {012315}
  (\bibinfo {year} {2018}{\natexlab{b}})}\BibitemShut {NoStop}%
\bibitem [{\citenamefont {Goldt}\ \emph {et~al.}(2019)\citenamefont {Goldt},
  \citenamefont {M\'ezard}, \citenamefont {Krzakala},\ and\ \citenamefont
  {Zdeborov\'a}}]{Goldt:2019}%
  \BibitemOpen
  \bibfield  {author} {\bibinfo {author} {\bibfnamefont {S.}~\bibnamefont
  {Goldt}}, \bibinfo {author} {\bibfnamefont {M.}~\bibnamefont {M\'ezard}},
  \bibinfo {author} {\bibfnamefont {F.}~\bibnamefont {Krzakala}}, \ and\
  \bibinfo {author} {\bibfnamefont {L.}~\bibnamefont {Zdeborov\'a}},\
  }\href@noop {} {\bibfield  {journal} {\bibinfo  {journal} {arXiv:1909.11500
  [stat.ML]}\ } (\bibinfo {year} {2019})}\BibitemShut {NoStop}%
\bibitem [{\citenamefont {Gerace}\ \emph {et~al.}(2020)\citenamefont {Gerace},
  \citenamefont {Loureiro}, \citenamefont {Krzakala}, \citenamefont
  {M{\'e}zard},\ and\ \citenamefont
  {Zdeborov{\'a}}}]{gerace2020generalisation}%
  \BibitemOpen
  \bibfield  {author} {\bibinfo {author} {\bibfnamefont {F.}~\bibnamefont
  {Gerace}}, \bibinfo {author} {\bibfnamefont {B.}~\bibnamefont {Loureiro}},
  \bibinfo {author} {\bibfnamefont {F.}~\bibnamefont {Krzakala}}, \bibinfo
  {author} {\bibfnamefont {M.}~\bibnamefont {M{\'e}zard}}, \ and\ \bibinfo
  {author} {\bibfnamefont {L.}~\bibnamefont {Zdeborov{\'a}}},\ }\href@noop {}
  {\bibfield  {journal} {\bibinfo  {journal} {arXiv preprint arXiv:2002.09339}\
  } (\bibinfo {year} {2020})}\BibitemShut {NoStop}%
\bibitem [{\citenamefont {Erba}\ \emph {et~al.}(2019)\citenamefont {Erba},
  \citenamefont {Gherardi},\ and\ \citenamefont {Rotondo}}]{Erba2019}%
  \BibitemOpen
  \bibfield  {author} {\bibinfo {author} {\bibfnamefont {V.}~\bibnamefont
  {Erba}}, \bibinfo {author} {\bibfnamefont {M.}~\bibnamefont {Gherardi}}, \
  and\ \bibinfo {author} {\bibfnamefont {P.}~\bibnamefont {Rotondo}},\ }\href
  {\doibase 10.1038/s41598-019-53549-9} {\bibfield  {journal} {\bibinfo
  {journal} {Scientific Reports}\ }\textbf {\bibinfo {volume} {9}},\ \bibinfo
  {pages} {17133} (\bibinfo {year} {2019})}\BibitemShut {NoStop}%
\bibitem [{\citenamefont {Ansuini}\ \emph {et~al.}(2019)\citenamefont
  {Ansuini}, \citenamefont {Laio}, \citenamefont {Macke},\ and\ \citenamefont
  {Zoccolan}}]{AnsuiniLaio:2019}%
  \BibitemOpen
  \bibfield  {author} {\bibinfo {author} {\bibfnamefont {A.}~\bibnamefont
  {Ansuini}}, \bibinfo {author} {\bibfnamefont {A.}~\bibnamefont {Laio}},
  \bibinfo {author} {\bibfnamefont {J.~H.}\ \bibnamefont {Macke}}, \ and\
  \bibinfo {author} {\bibfnamefont {D.}~\bibnamefont {Zoccolan}},\ }\href@noop
  {} {\enquote {\bibinfo {title} {Intrinsic dimension of data representations
  in deep neural networks},}\ } (\bibinfo {year} {2019}),\ \Eprint
  {http://arxiv.org/abs/1905.12784} {arXiv:1905.12784 [cs.LG]} \BibitemShut
  {NoStop}%
\bibitem [{\citenamefont {Facco}\ \emph {et~al.}(2017)\citenamefont {Facco},
  \citenamefont {d'Errico}, \citenamefont {Rodriguez},\ and\ \citenamefont
  {Laio}}]{2017FaccoRodriguezEtAlEstimatingTheIntrinsicDimensionOfDatasetsByAMinimalNeighborhoodInformation}%
  \BibitemOpen
  \bibfield  {author} {\bibinfo {author} {\bibfnamefont {E.}~\bibnamefont
  {Facco}}, \bibinfo {author} {\bibfnamefont {M.}~\bibnamefont {d'Errico}},
  \bibinfo {author} {\bibfnamefont {A.}~\bibnamefont {Rodriguez}}, \ and\
  \bibinfo {author} {\bibfnamefont {A.}~\bibnamefont {Laio}},\ }\href {\doibase
  10.1038/s41598-017-11873-y} {\bibfield  {journal} {\bibinfo  {journal}
  {Scientific Reports}\ }\textbf {\bibinfo {volume} {7}} (\bibinfo {year}
  {2017}),\ 10.1038/s41598-017-11873-y}\BibitemShut {NoStop}%
\bibitem [{\citenamefont {Erba}\ \emph {et~al.}(2020)\citenamefont {Erba},
  \citenamefont {Ariosto}, \citenamefont {Gherardi},\ and\ \citenamefont
  {Rotondo}}]{erba2020random}%
  \BibitemOpen
  \bibfield  {author} {\bibinfo {author} {\bibfnamefont {V.}~\bibnamefont
  {Erba}}, \bibinfo {author} {\bibfnamefont {S.}~\bibnamefont {Ariosto}},
  \bibinfo {author} {\bibfnamefont {M.}~\bibnamefont {Gherardi}}, \ and\
  \bibinfo {author} {\bibfnamefont {P.}~\bibnamefont {Rotondo}},\ }\href@noop
  {} {\bibfield  {journal} {\bibinfo  {journal} {arXiv preprint
  arXiv:2002.12272}\ } (\bibinfo {year} {2020})}\BibitemShut {NoStop}%
\bibitem [{\citenamefont {Vapnik}(2013)}]{vapnik2013nature}%
  \BibitemOpen
  \bibfield  {author} {\bibinfo {author} {\bibfnamefont {V.}~\bibnamefont
  {Vapnik}},\ }\href@noop {} {\emph {\bibinfo {title} {The nature of
  statistical learning theory}}}\ (\bibinfo  {publisher} {Springer science \&
  business media},\ \bibinfo {year} {2013})\BibitemShut {NoStop}%
\bibitem [{\citenamefont {Bousquet}\ \emph {et~al.}(2004)\citenamefont
  {Bousquet}, \citenamefont {Boucheron},\ and\ \citenamefont
  {Lugosi}}]{Bousquet2004}%
  \BibitemOpen
  \bibfield  {author} {\bibinfo {author} {\bibfnamefont {O.}~\bibnamefont
  {Bousquet}}, \bibinfo {author} {\bibfnamefont {S.}~\bibnamefont {Boucheron}},
  \ and\ \bibinfo {author} {\bibfnamefont {G.}~\bibnamefont {Lugosi}},\
  }\enquote {\bibinfo {title} {Introduction to statistical learning theory},}\
  in\ \href {\doibase 10.1007/978-3-540-28650-9_8} {\emph {\bibinfo {booktitle}
  {Advanced Lectures on Machine Learning: ML Summer Schools 2003, Canberra,
  Australia, February 2 - 14, 2003, T{\"u}bingen, Germany, August 4 - 16, 2003,
  Revised Lectures}}},\ \bibinfo {editor} {edited by\ \bibinfo {editor}
  {\bibfnamefont {O.}~\bibnamefont {Bousquet}}, \bibinfo {editor}
  {\bibfnamefont {U.}~\bibnamefont {von Luxburg}}, \ and\ \bibinfo {editor}
  {\bibfnamefont {G.}~\bibnamefont {R{\"a}tsch}}}\ (\bibinfo  {publisher}
  {Springer Berlin Heidelberg},\ \bibinfo {address} {Berlin, Heidelberg},\
  \bibinfo {year} {2004})\ pp.\ \bibinfo {pages} {169--207}\BibitemShut
  {NoStop}%
\bibitem [{\citenamefont {Zhang}\ \emph {et~al.}(2017)\citenamefont {Zhang},
  \citenamefont {Bengio}, \citenamefont {Hardt}, \citenamefont {Recht},\ and\
  \citenamefont {Vinyals}}]{ZhangBengio2017}%
  \BibitemOpen
  \bibfield  {author} {\bibinfo {author} {\bibfnamefont {C.}~\bibnamefont
  {Zhang}}, \bibinfo {author} {\bibfnamefont {S.}~\bibnamefont {Bengio}},
  \bibinfo {author} {\bibfnamefont {M.}~\bibnamefont {Hardt}}, \bibinfo
  {author} {\bibfnamefont {B.}~\bibnamefont {Recht}}, \ and\ \bibinfo {author}
  {\bibfnamefont {O.}~\bibnamefont {Vinyals}},\ }in\ \href
  {https://arxiv.org/abs/1611.03530} {\emph {\bibinfo {booktitle} {Proceedings
  of the International Conference on Learning Representations}}}\ (\bibinfo
  {year} {2017})\BibitemShut {NoStop}%
\bibitem [{\citenamefont {Martin}\ and\ \citenamefont
  {Mahoney}(2017)}]{MartinMahoney2017}%
  \BibitemOpen
  \bibfield  {author} {\bibinfo {author} {\bibfnamefont {C.~H.}\ \bibnamefont
  {Martin}}\ and\ \bibinfo {author} {\bibfnamefont {M.~W.}\ \bibnamefont
  {Mahoney}},\ }\href@noop {} {\bibfield  {journal} {\bibinfo  {journal}
  {arXiv:1710.09553 [cs.LG]}\ } (\bibinfo {year} {2017})}\BibitemShut {NoStop}%
\bibitem [{\citenamefont {Bottou}(2015)}]{Bottou2015}%
  \BibitemOpen
  \bibfield  {author} {\bibinfo {author} {\bibfnamefont {L.}~\bibnamefont
  {Bottou}},\ }\enquote {\bibinfo {title} {Making {V}apnik--{C}hervonenkis
  bounds accurate},}\ \ (\bibinfo {year} {2015})\ pp.\ \bibinfo {pages}
  {143--155}\BibitemShut {NoStop}%
\bibitem [{\citenamefont {{Cohn}}\ and\ \citenamefont
  {{Tesauro}}(1992)}]{Cohn1992}%
  \BibitemOpen
  \bibfield  {author} {\bibinfo {author} {\bibfnamefont {D.}~\bibnamefont
  {{Cohn}}}\ and\ \bibinfo {author} {\bibfnamefont {G.}~\bibnamefont
  {{Tesauro}}},\ }\href {\doibase 10.1162/neco.1992.4.2.249} {\bibfield
  {journal} {\bibinfo  {journal} {Neural Computation}\ }\textbf {\bibinfo
  {volume} {4}},\ \bibinfo {pages} {249} (\bibinfo {year} {1992})}\BibitemShut
  {NoStop}%
\bibitem [{\citenamefont {Antos}\ \emph {et~al.}(2003)\citenamefont {Antos},
  \citenamefont {K\'{e}gl}, \citenamefont {Linder},\ and\ \citenamefont
  {Lugosi}}]{Antos2003}%
  \BibitemOpen
  \bibfield  {author} {\bibinfo {author} {\bibfnamefont {A.}~\bibnamefont
  {Antos}}, \bibinfo {author} {\bibfnamefont {B.}~\bibnamefont {K\'{e}gl}},
  \bibinfo {author} {\bibfnamefont {T.}~\bibnamefont {Linder}}, \ and\ \bibinfo
  {author} {\bibfnamefont {G.}~\bibnamefont {Lugosi}},\ }\href {\doibase
  10.1162/153244303768966111} {\bibfield  {journal} {\bibinfo  {journal} {J.
  Mach. Learn. Res.}\ }\textbf {\bibinfo {volume} {3}},\ \bibinfo {pages} {73}
  (\bibinfo {year} {2003})}\BibitemShut {NoStop}%
\bibitem [{\citenamefont {K{\'e}gl}\ \emph {et~al.}(2001)\citenamefont
  {K{\'e}gl}, \citenamefont {Linder},\ and\ \citenamefont
  {Lugosi}}]{Lugosi2001}%
  \BibitemOpen
  \bibfield  {author} {\bibinfo {author} {\bibfnamefont {B.}~\bibnamefont
  {K{\'e}gl}}, \bibinfo {author} {\bibfnamefont {T.}~\bibnamefont {Linder}}, \
  and\ \bibinfo {author} {\bibfnamefont {G.}~\bibnamefont {Lugosi}},\ }in\
  \href@noop {} {\emph {\bibinfo {booktitle} {Computational Learning
  Theory}}},\ \bibinfo {editor} {edited by\ \bibinfo {editor} {\bibfnamefont
  {D.}~\bibnamefont {Helmbold}}\ and\ \bibinfo {editor} {\bibfnamefont
  {B.}~\bibnamefont {Williamson}}}\ (\bibinfo  {publisher} {Springer Berlin
  Heidelberg},\ \bibinfo {address} {Berlin, Heidelberg},\ \bibinfo {year}
  {2001})\ pp.\ \bibinfo {pages} {368--384}\BibitemShut {NoStop}%
\bibitem [{\citenamefont {{Shawe-Taylor}}\ \emph {et~al.}(1998)\citenamefont
  {{Shawe-Taylor}}, \citenamefont {{Bartlett}}, \citenamefont {{Williamson}},\
  and\ \citenamefont {{Anthony}}}]{Shawe-Taylor1998}%
  \BibitemOpen
  \bibfield  {author} {\bibinfo {author} {\bibfnamefont {J.}~\bibnamefont
  {{Shawe-Taylor}}}, \bibinfo {author} {\bibfnamefont {P.~L.}\ \bibnamefont
  {{Bartlett}}}, \bibinfo {author} {\bibfnamefont {R.~C.}\ \bibnamefont
  {{Williamson}}}, \ and\ \bibinfo {author} {\bibfnamefont {M.}~\bibnamefont
  {{Anthony}}},\ }\href {\doibase 10.1109/18.705570} {\bibfield  {journal}
  {\bibinfo  {journal} {IEEE Transactions on Information Theory}\ }\textbf
  {\bibinfo {volume} {44}},\ \bibinfo {pages} {1926} (\bibinfo {year}
  {1998})}\BibitemShut {NoStop}%
\bibitem [{\citenamefont {{Cover}}(1965)}]{cover1965}%
  \BibitemOpen
  \bibfield  {author} {\bibinfo {author} {\bibfnamefont {T.~M.}\ \bibnamefont
  {{Cover}}},\ }\href {\doibase 10.1109/PGEC.1965.264137} {\bibfield  {journal}
  {\bibinfo  {journal} {IEEE Transactions on Electronic Computers}\ }\textbf
  {\bibinfo {volume} {EC-14}},\ \bibinfo {pages} {326} (\bibinfo {year}
  {1965})}\BibitemShut {NoStop}%
\bibitem [{\citenamefont {{Vapnik}}(1999)}]{788640}%
  \BibitemOpen
  \bibfield  {author} {\bibinfo {author} {\bibfnamefont {V.~N.}\ \bibnamefont
  {{Vapnik}}},\ }\href {\doibase 10.1109/72.788640} {\bibfield  {journal}
  {\bibinfo  {journal} {IEEE Transactions on Neural Networks}\ }\textbf
  {\bibinfo {volume} {10}},\ \bibinfo {pages} {988} (\bibinfo {year}
  {1999})}\BibitemShut {NoStop}%
\bibitem [{\citenamefont {McCoy}(2010)}]{McCoy:BOOK}%
  \BibitemOpen
  \bibfield  {author} {\bibinfo {author} {\bibfnamefont {B.~M.}\ \bibnamefont
  {McCoy}},\ }\href@noop {} {\emph {\bibinfo {title} {Advanced statistical
  mechanics}}}\ (\bibinfo  {publisher} {Oxford University Press},\ \bibinfo
  {address} {Oxford},\ \bibinfo {year} {2010})\BibitemShut {NoStop}%
\bibitem [{\citenamefont {Caracciolo}\ \emph {et~al.}(2018)\citenamefont
  {Caracciolo}, \citenamefont {Di~Gioacchino}, \citenamefont {Gherardi},\ and\
  \citenamefont {Malatesta}}]{CaraccioloDiGioacchino:2018}%
  \BibitemOpen
  \bibfield  {author} {\bibinfo {author} {\bibfnamefont {S.}~\bibnamefont
  {Caracciolo}}, \bibinfo {author} {\bibfnamefont {A.}~\bibnamefont
  {Di~Gioacchino}}, \bibinfo {author} {\bibfnamefont {M.}~\bibnamefont
  {Gherardi}}, \ and\ \bibinfo {author} {\bibfnamefont {E.~M.}\ \bibnamefont
  {Malatesta}},\ }\href {\doibase 10.1103/PhysRevE.97.052109} {\bibfield
  {journal} {\bibinfo  {journal} {Phys. Rev. E}\ }\textbf {\bibinfo {volume}
  {97}},\ \bibinfo {pages} {052109} (\bibinfo {year} {2018})}\BibitemShut
  {NoStop}%
\bibitem [{\citenamefont {Caracciolo}\ and\ \citenamefont
  {Sportiello}(2002)}]{CaraccioloSportiello:2002}%
  \BibitemOpen
  \bibfield  {author} {\bibinfo {author} {\bibfnamefont {S.}~\bibnamefont
  {Caracciolo}}\ and\ \bibinfo {author} {\bibfnamefont {A.}~\bibnamefont
  {Sportiello}},\ }\href {\doibase 10.1088/0305-4470/35/36/301} {\bibfield
  {journal} {\bibinfo  {journal} {Journal of Physics A: Mathematical and
  General}\ }\textbf {\bibinfo {volume} {35}},\ \bibinfo {pages} {7661}
  (\bibinfo {year} {2002})}\BibitemShut {NoStop}%
\bibitem [{\citenamefont {Rotondo}\ \emph {et~al.}(2020)\citenamefont
  {Rotondo}, \citenamefont {Lagomarsino},\ and\ \citenamefont
  {Gherardi}}]{Rotondo:2020:PRR}%
  \BibitemOpen
  \bibfield  {author} {\bibinfo {author} {\bibfnamefont {P.}~\bibnamefont
  {Rotondo}}, \bibinfo {author} {\bibfnamefont {M.~C.}\ \bibnamefont
  {Lagomarsino}}, \ and\ \bibinfo {author} {\bibfnamefont {M.}~\bibnamefont
  {Gherardi}},\ }\href {\doibase 10.1103/PhysRevResearch.2.023169} {\bibfield
  {journal} {\bibinfo  {journal} {Phys. Rev. Research}\ }\textbf {\bibinfo
  {volume} {2}},\ \bibinfo {pages} {023169} (\bibinfo {year}
  {2020})}\BibitemShut {NoStop}%
\bibitem [{\citenamefont {Cortes}\ and\ \citenamefont
  {Vapnik}(1995)}]{Cortes1995}%
  \BibitemOpen
  \bibfield  {author} {\bibinfo {author} {\bibfnamefont {C.}~\bibnamefont
  {Cortes}}\ and\ \bibinfo {author} {\bibfnamefont {V.}~\bibnamefont
  {Vapnik}},\ }\href {\doibase 10.1007/BF00994018} {\bibfield  {journal}
  {\bibinfo  {journal} {Machine Learning}\ }\textbf {\bibinfo {volume} {20}},\
  \bibinfo {pages} {273} (\bibinfo {year} {1995})}\BibitemShut {NoStop}%
\bibitem [{\citenamefont {and Rotondo}\ \emph {et~al.}()\citenamefont {and
  Rotondo}, \citenamefont {Pastore},\ and\ \citenamefont
  {Gherardi}}]{companion1}%
  \BibitemOpen
  \bibfield  {author} {\bibinfo {author} {\bibfnamefont {P.}~\bibnamefont {and
  Rotondo}}, \bibinfo {author} {\bibfnamefont {M.}~\bibnamefont {Pastore}}, \
  and\ \bibinfo {author} {\bibfnamefont {M.}~\bibnamefont {Gherardi}},\
  }\href@noop {} {\bibinfo  {journal} {[in preparation]}\ }\BibitemShut
  {NoStop}%
\bibitem [{\citenamefont {Mehta}\ \emph {et~al.}(2019)\citenamefont {Mehta},
  \citenamefont {Bukov}, \citenamefont {Wang}, \citenamefont {Day},
  \citenamefont {Richardson}, \citenamefont {Fisher},\ and\ \citenamefont
  {Schwab}}]{Mehta:2019:PR}%
  \BibitemOpen
\bibfield  {journal} {  }\bibfield  {author} {\bibinfo {author} {\bibfnamefont
  {P.}~\bibnamefont {Mehta}}, \bibinfo {author} {\bibfnamefont
  {M.}~\bibnamefont {Bukov}}, \bibinfo {author} {\bibfnamefont {C.-H.}\
  \bibnamefont {Wang}}, \bibinfo {author} {\bibfnamefont {A.~G.}\ \bibnamefont
  {Day}}, \bibinfo {author} {\bibfnamefont {C.}~\bibnamefont {Richardson}},
  \bibinfo {author} {\bibfnamefont {C.~K.}\ \bibnamefont {Fisher}}, \ and\
  \bibinfo {author} {\bibfnamefont {D.~J.}\ \bibnamefont {Schwab}},\ }\href
  {\doibase https://doi.org/10.1016/j.physrep.2019.03.001} {\bibfield
  {journal} {\bibinfo  {journal} {Physics Reports}\ }\textbf {\bibinfo {volume}
  {810}},\ \bibinfo {pages} {1 } (\bibinfo {year} {2019})},\ \bibinfo {note} {a
  high-bias, low-variance introduction to Machine Learning for
  physicists}\BibitemShut {NoStop}%
\bibitem [{\citenamefont {Sontag}(1998)}]{sontag1998vc}%
  \BibitemOpen
  \bibfield  {author} {\bibinfo {author} {\bibfnamefont {E.~D.}\ \bibnamefont
  {Sontag}},\ }\href@noop {} {\bibfield  {journal} {\bibinfo  {journal} {NATO
  ASI Series F Computer and Systems Sciences}\ }\textbf {\bibinfo {volume}
  {168}},\ \bibinfo {pages} {69} (\bibinfo {year} {1998})}\BibitemShut
  {NoStop}%
\bibitem [{\citenamefont {Bartlett}\ and\ \citenamefont
  {Mendelson}(2003)}]{BartlettMendelson:2003}%
  \BibitemOpen
  \bibfield  {author} {\bibinfo {author} {\bibfnamefont {P.~L.}\ \bibnamefont
  {Bartlett}}\ and\ \bibinfo {author} {\bibfnamefont {S.}~\bibnamefont
  {Mendelson}},\ }\href@noop {} {\bibfield  {journal} {\bibinfo  {journal} {J.
  Mach. Learn. Res.}\ }\textbf {\bibinfo {volume} {3}},\ \bibinfo {pages} {463}
  (\bibinfo {year} {2003})}\BibitemShut {NoStop}%
\bibitem [{\citenamefont {Abbara}\ \emph {et~al.}(2019)\citenamefont {Abbara},
  \citenamefont {Aubin}, \citenamefont {Krzakala},\ and\ \citenamefont
  {Zdeborov{\'a}}}]{AbbaraAubin:2019}%
  \BibitemOpen
  \bibfield  {author} {\bibinfo {author} {\bibfnamefont {A.}~\bibnamefont
  {Abbara}}, \bibinfo {author} {\bibfnamefont {B.}~\bibnamefont {Aubin}},
  \bibinfo {author} {\bibfnamefont {F.}~\bibnamefont {Krzakala}}, \ and\
  \bibinfo {author} {\bibfnamefont {L.}~\bibnamefont {Zdeborov{\'a}}},\
  }\href@noop {} {\enquote {\bibinfo {title} {Rademacher complexity and spin
  glasses: A link between the replica and statistical theories of learning},}\
  } (\bibinfo {year} {2019}),\ \Eprint {http://arxiv.org/abs/1912.02729}
  {arXiv:1912.02729 [cond-mat.dis-nn]} \BibitemShut {NoStop}%
\bibitem [{\citenamefont {{Anguita}}\ \emph {et~al.}(2014)\citenamefont
  {{Anguita}}, \citenamefont {{Ghio}}, \citenamefont {{Oneto}},\ and\
  \citenamefont {{Ridella}}}]{AnguitaGhio:2014}%
  \BibitemOpen
  \bibfield  {author} {\bibinfo {author} {\bibfnamefont {D.}~\bibnamefont
  {{Anguita}}}, \bibinfo {author} {\bibfnamefont {A.}~\bibnamefont {{Ghio}}},
  \bibinfo {author} {\bibfnamefont {L.}~\bibnamefont {{Oneto}}}, \ and\
  \bibinfo {author} {\bibfnamefont {S.}~\bibnamefont {{Ridella}}},\ }\href
  {\doibase 10.1109/TNNLS.2014.2307359} {\bibfield  {journal} {\bibinfo
  {journal} {IEEE Transactions on Neural Networks and Learning Systems}\
  }\textbf {\bibinfo {volume} {25}},\ \bibinfo {pages} {2202} (\bibinfo {year}
  {2014})}\BibitemShut {NoStop}%
\bibitem [{\citenamefont {Huang}\ \emph {et~al.}(2006)\citenamefont {Huang},
  \citenamefont {Zhu},\ and\ \citenamefont {Siew}}]{HUANG2006489}%
  \BibitemOpen
  \bibfield  {author} {\bibinfo {author} {\bibfnamefont {G.-B.}\ \bibnamefont
  {Huang}}, \bibinfo {author} {\bibfnamefont {Q.-Y.}\ \bibnamefont {Zhu}}, \
  and\ \bibinfo {author} {\bibfnamefont {C.-K.}\ \bibnamefont {Siew}},\ }\href
  {\doibase https://doi.org/10.1016/j.neucom.2005.12.126} {\bibfield  {journal}
  {\bibinfo  {journal} {Neurocomputing}\ }\textbf {\bibinfo {volume} {70}},\
  \bibinfo {pages} {489 } (\bibinfo {year} {2006})},\ \bibinfo {note} {neural
  Networks}\BibitemShut {NoStop}%
\bibitem [{\citenamefont {Anselmi}\ \emph {et~al.}(2016)\citenamefont
  {Anselmi}, \citenamefont {Leibo}, \citenamefont {Rosasco}, \citenamefont
  {Mutch}, \citenamefont {Tacchetti},\ and\ \citenamefont
  {Poggio}}]{AnselmiLeibo:2016}%
  \BibitemOpen
  \bibfield  {author} {\bibinfo {author} {\bibfnamefont {F.}~\bibnamefont
  {Anselmi}}, \bibinfo {author} {\bibfnamefont {J.~Z.}\ \bibnamefont {Leibo}},
  \bibinfo {author} {\bibfnamefont {L.}~\bibnamefont {Rosasco}}, \bibinfo
  {author} {\bibfnamefont {J.}~\bibnamefont {Mutch}}, \bibinfo {author}
  {\bibfnamefont {A.}~\bibnamefont {Tacchetti}}, \ and\ \bibinfo {author}
  {\bibfnamefont {T.}~\bibnamefont {Poggio}},\ }\href {\doibase
  10.1016/j.tcs.2015.06.048} {\bibfield  {journal} {\bibinfo  {journal} {Theor.
  Comput. Sci.}\ }\textbf {\bibinfo {volume} {633}},\ \bibinfo {pages} {112}
  (\bibinfo {year} {2016})}\BibitemShut {NoStop}%
\bibitem [{\citenamefont {Seung}\ and\ \citenamefont {Lee}(2000)}]{Seung:2000}%
  \BibitemOpen
  \bibfield  {author} {\bibinfo {author} {\bibfnamefont {H.~S.}\ \bibnamefont
  {Seung}}\ and\ \bibinfo {author} {\bibfnamefont {D.~D.}\ \bibnamefont
  {Lee}},\ }\href {\doibase 10.1126/science.290.5500.2268} {\bibfield
  {journal} {\bibinfo  {journal} {Science}\ }\textbf {\bibinfo {volume}
  {290}},\ \bibinfo {pages} {2268} (\bibinfo {year} {2000})}\BibitemShut
  {NoStop}%
\bibitem [{\citenamefont {Borra}\ \emph {et~al.}(2019)\citenamefont {Borra},
  \citenamefont {Lagomarsino}, \citenamefont {Rotondo},\ and\ \citenamefont
  {Gherardi}}]{Borra:2019}%
  \BibitemOpen
  \bibfield  {author} {\bibinfo {author} {\bibfnamefont {F.}~\bibnamefont
  {Borra}}, \bibinfo {author} {\bibfnamefont {M.~C.}\ \bibnamefont
  {Lagomarsino}}, \bibinfo {author} {\bibfnamefont {P.}~\bibnamefont
  {Rotondo}}, \ and\ \bibinfo {author} {\bibfnamefont {M.}~\bibnamefont
  {Gherardi}},\ }\href {\doibase 10.1088/1751-8121/ab3709} {\bibfield
  {journal} {\bibinfo  {journal} {Journal of Physics A: Mathematical and
  Theoretical}\ }\textbf {\bibinfo {volume} {52}},\ \bibinfo {pages} {384004}
  (\bibinfo {year} {2019})}\BibitemShut {NoStop}%
\bibitem [{\citenamefont {Flajolet}\ and\ \citenamefont
  {Sedgewick}(2009)}]{flajolet2009analytic}%
  \BibitemOpen
  \bibfield  {author} {\bibinfo {author} {\bibfnamefont {P.}~\bibnamefont
  {Flajolet}}\ and\ \bibinfo {author} {\bibfnamefont {R.}~\bibnamefont
  {Sedgewick}},\ }\href@noop {} {\emph {\bibinfo {title} {Analytic
  combinatorics}}}\ (\bibinfo  {publisher} {cambridge University press},\
  \bibinfo {year} {2009})\BibitemShut {NoStop}%
\bibitem [{\citenamefont {Kirkpatrick}\ and\ \citenamefont
  {Selman}(1994)}]{Kirkpatrick1994}%
  \BibitemOpen
  \bibfield  {author} {\bibinfo {author} {\bibfnamefont {S.}~\bibnamefont
  {Kirkpatrick}}\ and\ \bibinfo {author} {\bibfnamefont {B.}~\bibnamefont
  {Selman}},\ }\href {\doibase 10.1126/science.264.5163.1297} {\bibfield
  {journal} {\bibinfo  {journal} {Science}\ }\textbf {\bibinfo {volume}
  {264}},\ \bibinfo {pages} {1297} (\bibinfo {year} {1994})}\BibitemShut
  {NoStop}%
\bibitem [{\citenamefont {Leone}\ \emph {et~al.}(2001)\citenamefont {Leone},
  \citenamefont {Ricci-Tersenghi},\ and\ \citenamefont {Zecchina}}]{Leone2001}%
  \BibitemOpen
  \bibfield  {author} {\bibinfo {author} {\bibfnamefont {M.}~\bibnamefont
  {Leone}}, \bibinfo {author} {\bibfnamefont {F.}~\bibnamefont
  {Ricci-Tersenghi}}, \ and\ \bibinfo {author} {\bibfnamefont {R.}~\bibnamefont
  {Zecchina}},\ }\href {\doibase 10.1088/0305-4470/34/22/303} {\bibfield
  {journal} {\bibinfo  {journal} {Journal of Physics A: Mathematical and
  General}\ }\textbf {\bibinfo {volume} {34}},\ \bibinfo {pages} {4615}
  (\bibinfo {year} {2001})}\BibitemShut {NoStop}%
\bibitem [{\citenamefont {Gupta}(1963)}]{gupta1963}%
  \BibitemOpen
  \bibfield  {author} {\bibinfo {author} {\bibfnamefont {S.~S.}\ \bibnamefont
  {Gupta}},\ }\href {\doibase 10.1214/aoms/1177704004} {\bibfield  {journal}
  {\bibinfo  {journal} {Ann. Math. Statist.}\ }\textbf {\bibinfo {volume}
  {34}},\ \bibinfo {pages} {792} (\bibinfo {year} {1963})}\BibitemShut
  {NoStop}%
\bibitem [{\citenamefont {Baldassi}\ \emph {et~al.}(2019)\citenamefont
  {Baldassi}, \citenamefont {Malatesta},\ and\ \citenamefont
  {Zecchina}}]{Malatesta_2019}%
  \BibitemOpen
  \bibfield  {author} {\bibinfo {author} {\bibfnamefont {C.}~\bibnamefont
  {Baldassi}}, \bibinfo {author} {\bibfnamefont {E.~M.}\ \bibnamefont
  {Malatesta}}, \ and\ \bibinfo {author} {\bibfnamefont {R.}~\bibnamefont
  {Zecchina}},\ }\href {\doibase 10.1103/PhysRevLett.123.170602} {\bibfield
  {journal} {\bibinfo  {journal} {Phys. Rev. Lett.}\ }\textbf {\bibinfo
  {volume} {123}},\ \bibinfo {pages} {170602} (\bibinfo {year}
  {2019})}\BibitemShut {NoStop}%
\end{thebibliography}%

\end{document}